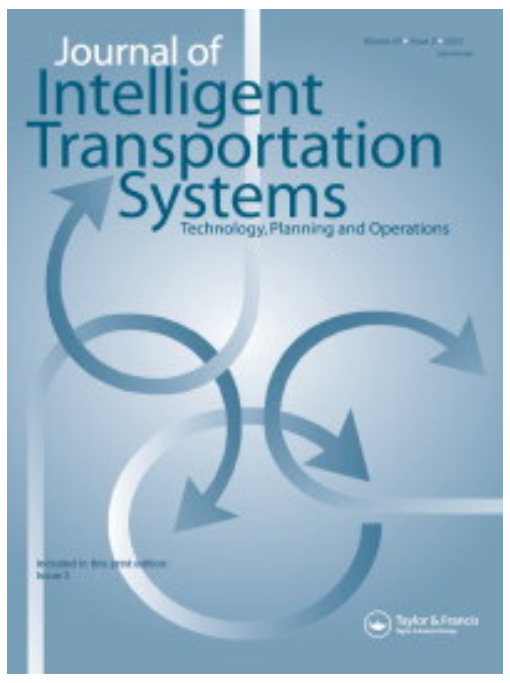

# Vehicle occupancy estimation in Automated Guideway Transit *via* deep learning with Wi-Fi probe requests

Ziyue Li & Qianwen Guo

# Vehicle occupancy estimation in Automated Guideway Transit *via* deep learning with Wi-Fi probe requests

Ziyue Li and Qianwen Guo

Department of Civil and Environmental Engineering, FAMU-FSU College of Engineering, Florida State University, Tallahassee, FL, USA

**ABSTRACT**

This study contributes to the advancement of vehicle occupancy estimation in Automated Guideway Transit (AGT) systems using Wi-Fi probe requests and deep learning models. We propose a comprehensive framework for evaluating various approaches to occupancy estimation, particularly in the context of MAC address randomization. While many methods proposed in the literature claim effectiveness in simpler experimental settings, our research reveals that those methods are unreliable in the complex environment of AGT systems. Specifically, techniques for handling randomized MAC addresses and distinguishing between passenger and non-passenger data do not perform well in AGT systems. Despite challenges in tracking individual devices, our study demonstrates that accurate occupancy estimation using Wi-Fi probe requests remains feasible. A pilot study conducted on the Miami-Dade Metromover, an AGT system characterized by frequent stops, significant occupancy fluctuations, and absence of fare collection devices, provides a robust testing ground for the framework. Additionally, our findings show that deep learning models significantly outperform machine learning models in this context. The insights from this study can significantly enhance decision-making for transit agencies to optimize operations and elevate service quality.

## 1. Introduction

Transit occupancy information is imperative for transit planners and policy makers to deliver high-quality transit services, as a vehicle's load factor and crowding directly impact passenger comfort and well-being during travel (Maltinti et al., 2024). Based on in-vehicle occupancy, transit agencies can judiciously allocate resources (Nitti et al., 2020). This information also facilitates efficient trip planning for passengers (Mehmood et al., 2019). Despite the importance of occupancy data, transit agencies encounter challenges in its collection. Historically reliant on manual counting, they have increasingly adopted Automated Fare Collection (AFC) systems for occupancy estimation due to the advent of transit cards and ticket vending machines (Li et al., 2024; Lu et al., 2025). Technological advancements drive the evolution of Automatic Passenger Counting (APC) systems employing diverse technologies, such as infrared sensors, cameras, or weight sensors (Moore et al., 2002). Presently, with over 90% Americans using smartphones (Pew Research Center, 2024), opportunities for occupancy estimation have expanded through the utilization of Bluetooth (Kostakos et al., 2010) and Wi-Fi (Da Silva & Shalaby, 2024).

Occupancy estimation using Wi-Fi technology offers several advantages (Mikkelsen et al., 2016; Ryu et al., 2020). First, it is cost-effective and easy to install the device. Second, it can achieve real-time occupancy estimation when Wi-Fi frames are collected and processed timely. Third, it demonstrates strengths in privacy protection, as the designed algorithm does not require the physical device identifier.

As illustrated in Figure 1, the fundamental concept of estimating vehicle occupancy using Wi-Fi technology involves analyzing Wi-Fi packets generated by passengers' mobile devices while they are onboard (Mikkelsen et al., 2016). Even if not connected to a wireless network, Wi-Fi-enabled devices send out probe requests – periodic scans to search for available Wi-Fi networks by broadcasting packets (frames) (Freudiger, 2015). These probe requests can subsequently be captured by Wi-Fi sniffers. The probe request contains unencrypted information, including the Media Access Control (MAC) address, Received Signal Strength Indicator (RSSI),

**CONTACT** Qianwen Guo qguo@eng.famu.fsu.edu Department of Civil and Environmental Engineering, FAMU-FSU College of Engineering, Florida State University, Tallahassee, FL, USA

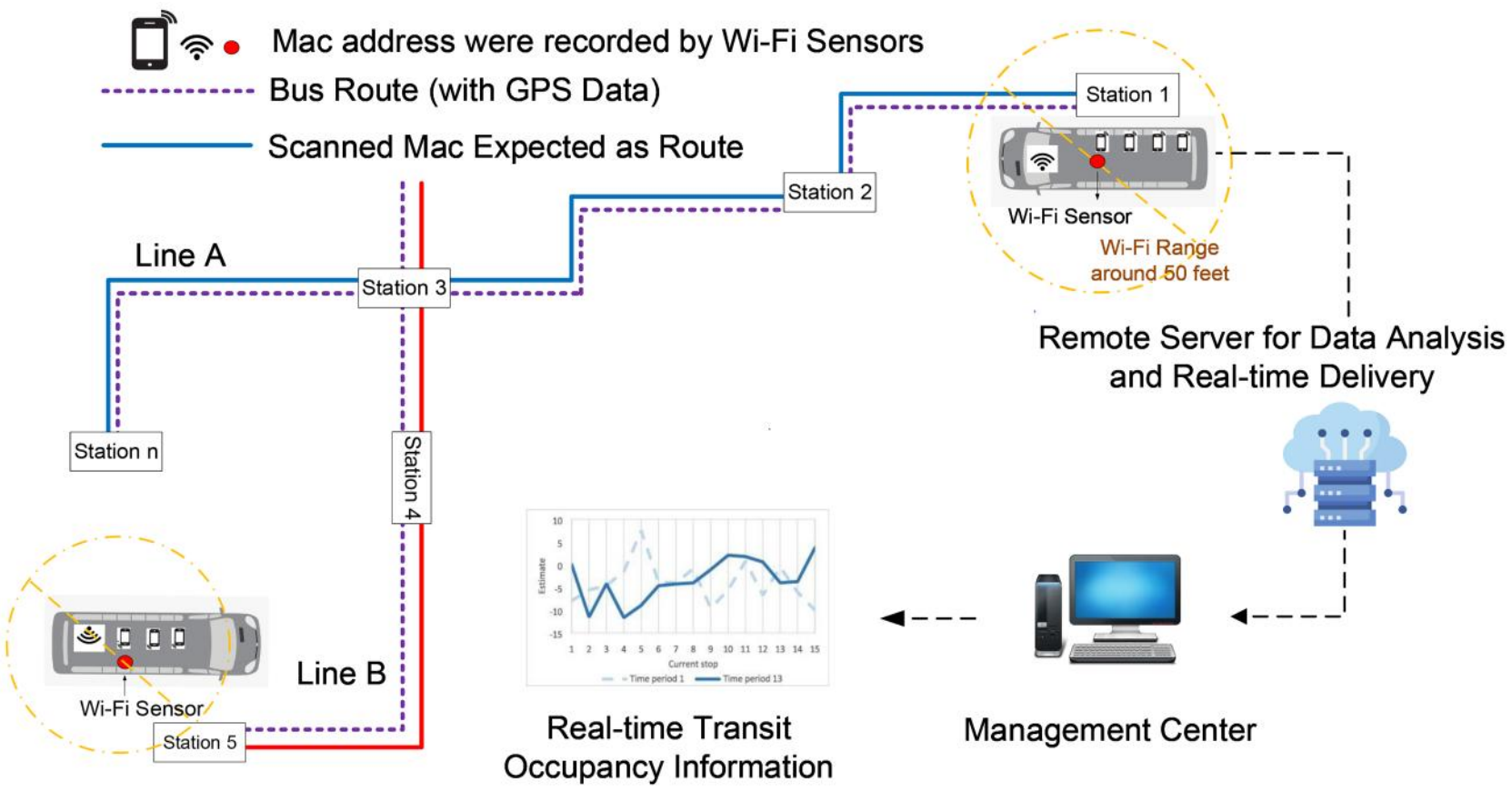

**Figure 1.** Wi-Fi Technology in public transit.

Service Set Identifier (SSID), and data rate, facilitating data analysis through a digital fingerprint (Pu et al., 2021).

A MAC address serves as a unique identifier assigned to a network interface for intra-network communications, enabling the distinction of individual devices (Pu et al., 2021). However, with the advancement of privacy protection measures, most smart devices now feature MAC address randomization enabled by default. This feature results in nearly every field in a probe request from a single device showing inconsistent values (Newell, 2023), presenting a substantial challenge in accurately identifying and tracking devices. Previous research utilizing Wi-Fi technology, while demonstrating satisfactory performance, often did not account for MAC address randomization (Oransirikul & Takada, 2019; Pu et al., 2021). Even recent studies focusing on inferring travel patterns, such as Gao and Schmöcker (2024), analyzed only a subset of devices with non-randomized MAC addresses. This oversight results in significant data underutilization and introduces potential biases. Another study, Fabre et al. (2024), collected data with true MAC addresses, claiming that "the MAC addresses are pseudonymized within very short delays and posters are displayed in the bus to warn passengers that some sensors are active" to comply with personal data protection laws. However, passengers might choose to disable their devices' Wi-Fi to protect privacy upon noticing the warning. Although various methods have been proposed to address the challenges posed by MAC address randomization, their effectiveness in real-world complex transit environments remains uncertain. For instance, Nitti et al. (2020) were among the first studies to analyze this issue. However, like other similar studies, its methods were only validated under controlled experimental conditions, leaving their applicability in complex real-world settings unverified. Moreover, specific de-randomization techniques can quickly become obsolete, as smart device manufacturers continuously update MAC address randomization protocols, and different manufacturers employ distinct randomization schemes (Fenske et al., 2021). Further discussion is provided in Section 2.2.

Additionally, previous studies were mostly conducted in experimental settings or in vehicles where both the range and maximum occupancy were relatively small. Real-world complex transit systems, such as the Automated Guideway Transit (AGT), present other distinctive challenges for occupancy estimation (Pastor, 1988). First, the travel time between two stops is relatively short, with vehicles in extreme cases passing through multiple stations within one minute. Therefore, the approach to data aggregation must be meticulously designed. Second, due to the system's fare-free operation (passengers can board and alight freely), AFC data are unavailable. Since AFC data, when combined with other data sources, can improve the accuracy and robustness of occupancy estimation (Dib et al., 2023), its absence requires careful handling to address potential over- or underestimation biases. Notably, no studies have yet explored the use of Wi-Fi technology to estimate occupancy within AGT systems. This study marks the first attempt to apply Wi-Fi technology for monitoring occupancy in AGT systems, which differ significantly from other transit modes.

This study aims to investigate the challenges of Wi-Fi-based occupancy estimation for AGT systems. Our contributions are outlined as follows:

1. A comprehensive framework for occupancy estimation using Wi-Fi is developed. We tackle prevalent obstacles, such as MAC addresses

randomization, differentiating between passengers and non-passengers, and the data aggregation process.
2. Deep learning is applied for occupancy estimation. This approach is relatively unexplored and it enhances the precision and reliability of occupancy estimation.
3. A pilot study in the Miami-Dade Metromover is conducted. The study navigates through the intricate challenges unique to AGT, setting a foundational precedent for future research and development in this domain.

Our research reveals a series of key findings:

1. There are no reliable density-based methods to handle randomized MAC addresses nor fuzzy clustering algorithms to classify passengers and non-passengers in AGT.
2. Our experiments reveal an important concept not extensively discussed in the literature: the burst. During each burst, the MAC address remains constant (Kumar & Cunche, 2024). Consequently, the number of bursts becomes an important feature for occupancy estimation. This may explain why, even without MAC address de-randomization, satisfactory occupancy estimation accuracy can still be achieved by leveraging this feature along with others.
3. A notable improvement in occupancy estimation accuracy is demonstrated by implementing deep learning techniques. It is found that treating vehicle occupancy as a time series and incorporating past data spanning 8–10 time lags (corresponding to 8–10 min) results in robust performance.

The remainder of this article is organized as follows. Section 2 reviews the relevant literature. Section 3 describes the data collection and pre-processing procedure. Section 4 discusses the detailed methodology and Section 5 shows the experimental results. Finally, conclusions are drawn in Section 6.

## 2. Literature review

### 2.1. Wi-Fi sniffing system applications in public transit

Wi-Fi sniffing system applications in public transit can be categorized into two groups based on the deployment scale: those implemented in systems covering a limited number of spots and those deployed across extensive networks encompassing numerous locations (Ryu et al., 2020).

Within the first group where the Wi-Fi sniffing system is deployed in a confined area, Chon et al. (2014) detected the movement of surrounding users in a room, while Schauer et al. (2014) estimated the spatial density of pedestrians at a German airport. Pronello et al. (2025) counted the number of waiting passengers at bus stops, while Mehmood et al. (2019); Myrvoll et al. (2017) counting estimated bus occupancy based on Wi-Fi signatures. Maiti and Chilukuri (2024) developed a methodology to use Wi-Fi sensors for traffic state characterization (e.g., speed, queue length, and queue location from the sensor) on urban roads.

For Wi-Fi sniffers deployed at multiple locations, studies based on MAC re-identification (i.e., matching MAC addresses of individuals' mobile devices from multiple spots) make it possible to track trips and measure travel information. Dunlap et al. (2016) and Fabre et al. (2023) have estimated origin-destination (OD) information from the passive data collected by the sniffers. Gao and Schmöcker (2024) combined Wi-Fi sensing data and GPS traces to infer the travel patterns and the attractiveness of touristic areas. tu2019vifi developed ViFi-MobiScanner, a system that infers user mobility from network activity using 4800 mobile routers deployed across a city. However, these studies did not account for MAC address randomization. Without access to the true MAC address, estimating OD pairs becomes nearly intractable.

Table 1 summarizes related studies using Wi-Fi technology and their limitations. Most early studies did not consider MAC address randomization, as it was not yet implemented by smartphone manufacturers at the time. Consequently, their methods are not effective in current environments. While some recent studies account for MAC address randomization, their applicability remains limited to controlled lab settings with low-density scenarios. Further discussion is provided in Section 2.2.

### 2.2. MAC address randomization and handling methods

A MAC address consists of 48 bits, typically represented in hexadecimal format (Martin et al., 2017). The first half (24 bits) of a MAC address, named the Organization Unique Identifier (OUI), is associated with the device manufacturer, while the second half is unique to the device itself. Key bits determine its characteristics. For instance, the least significant bit ($b_0$) of the first byte distinguishes between unicast (0) and multicast (1) addresses, with unicast addresses designated for one-to-one communication and multicast for one-to-many communication. The second least significant bit ($b_1$) indicates whether the address is globally administered (0) or locally administered (1). Specifically, when $b_0$ denotes unicast and $b_1$ signifies locally administered, it implies a randomized MAC address (see Figure 2). This delineation allows for the recognition of specific byte patterns, "2,6,A,E", derived from the hexadecimal representation of "??10", to identify randomized MAC addresses. It is worth noting that certain Android devices may deviate from these conventions, maintaining a fixed Google OUI (`DA:A1:19`) while randomizing the remaining portion of the MAC address (Fenske et al., 2021). Nonetheless, in this study, we assume that "2,6,A,E" represents randomized MAC addresses. Today, most mobile device manufacturers incorporate MAC address randomization to protect user privacy. This technique varies significantly across devices and operating systems (Fenske et al., 2021). Devices may change MAC addresses after each network connection or at regular intervals, presenting a substantial challenge for device identification.

**Table 1.** Existing studies using Wi-Fi technology and their limitations.

| Study | Problem | Consider MAC address randomization? |
|---|---|---|
| Chon et al. (2014) | Movement detection | No |
| Schauer et al. (2014) | Density estimation and flow detection | No |
| Pronello et al. (2025) | People counting | Yes, but in controlled settings with low density |
| Mehmood et al. (2019) | Vehicle occupancy estimation | No |
| Myrvoll et al. (2017) | Vehicle occupancy estimation | No |
| Dunlap et al. (2016) | Transit O-D estimation | No |
| Fabre et al. (2023) | Transit O-D estimation | No |
| Gao and Schmöcker (2024) | Tourist trip-chain inference | No |
| Tu et al. (2019) | Human mobility analysis | No |

In the realm of occupancy estimation utilizing Wi-Fi technology, there exists a body of research addressing the challenges posed by randomized MAC addresses (Delzanno et al., 2023; Newell, 2023; Pronello et al., 2025; Uras et al., 2020), among which the predominant technique is density-based clustering. Uras et al. (2020) pioneered the application of clustering algorithms, such as Density-Based Spatial Clustering of Application with Noise (DBSCAN) and Ordering Points to Identify the Clustering Structure (OPTICS), to analyze features for device identification. Despite achieving high accuracy, their experiments occurred within controlled laboratory settings, involving only 23 devices. Delzanno et al. (2023) designed a simplified DBSCAN algorithm, named $\text{DBSCAN}_\text{L}$, to cluster MAC addresses according to RSSI and data rate. Tan and Chan (2021) proposed the ESPRESSO algorithm, leveraging Bayes' theorem to establish the associative probability between bursts of probe requests. Brik et al. (2008) analyzed the physical layer and suggested a method by performing passive analysis of radio frequencies and employing machine learning to achieve a 99% accuracy in device counting, but it is impractical in real-world scenarios due to radio interference and the complexity of data collection setup. Matte et al. (2016) proposed a way to fingerprint the probe requests sent by a single device. Although they showed that such time-based signature is consistent over time, the methodology highly relied on the fact that "the frames sent by Wi-Fi devices follow regular patterns that can be used for time-based fingerprinting (Franklin et al., 2006)." Timing information is not reliable due to scattering problems and multi-path phenomena that occur in realistic settings (Uras et al., 2020).

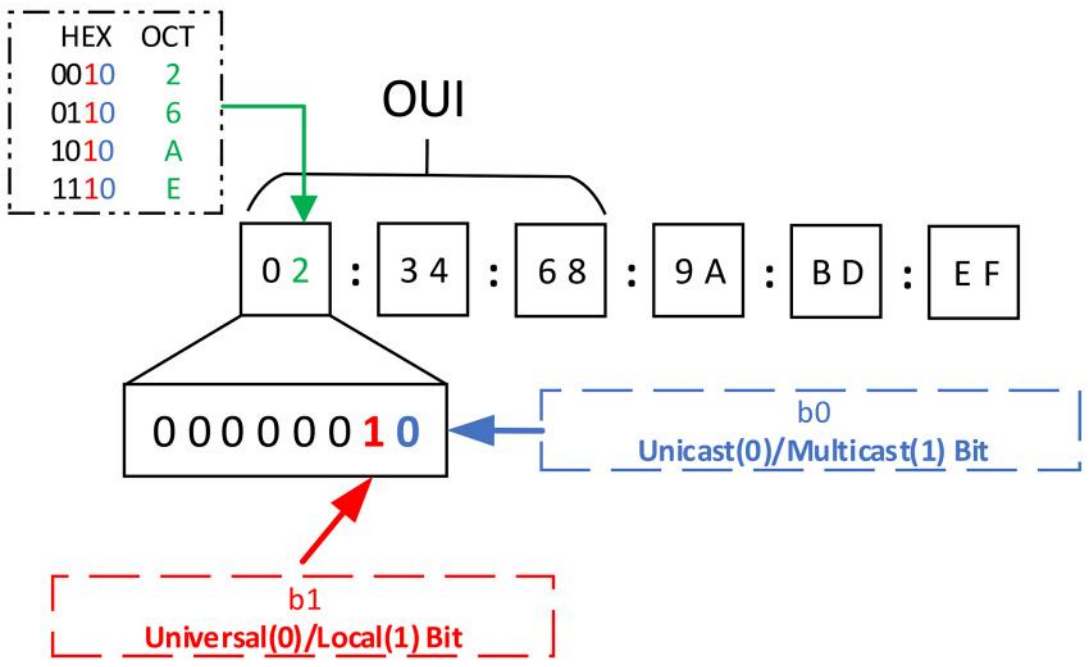

**Figure 2.** MAC address structure.

A summary of existing methods addressing MAC address randomization is provided in Table 2. These methods have only been validated in controlled lab settings, and their effectiveness in real-world AGT systems remains unknown. To the best of our knowledge, our study is the first to examine these challenges in such a complex environment. Figure 3 outlines our investigative approach, presenting key questions that guide our analysis.

## 2.3. Occupancy estimation using Wi-Fi technology

There are two Wi-Fi-based vehicle occupancy estimation approaches, depending on whether users are required to connect their devices to a specific Wi-Fi network. For the first method termed "active collecting," devices are required to be connected to a certain wireless network that covers the entire study area (Ozbay et al., 2017). Thus, the area under study cannot be expanded without extending the network coverage, which may be challenging in some cases (Lesani & Miranda-Moreno, 2019). In contrast, our focus lies on "passive sniffing" methods (Fabre et al., 2023; Mehmood et al., 2019; Mishalani et al., 2016). Compared to active collecting, passive sniffing offers superior privacy protection and does not require device connection to specific Wi-Fi networks. However, most passive sniffing studies rely on non-randomized MAC addresses emitted by Wi-Fi devices. The increasing adoption of MAC address randomization by smartphone manufacturers disrupts device identification accuracy.

**Table 2.** Methods to address MAC address randomization.

| Study | Method | Limitation |
|---|---|---|
| Nitti et al. (2020) | Cluster probe requests with matching identifier fields and assign a probability score to determine whether two probe requests originate from the same device | Validated only in controlled experimental settings with a maximum of 8 passengers and predefined devices |
| Uras et al. (2020) | Cluster probe requests based on similar frame features; DBSCAN and OPTICS algorithm | Effective only in not crowded environments; use the number of identified devices as a criterion rather than actual occupancy |
| Delzanno et al. (2023) | Cluster probe requests using RSSI and data rates; DBSCAN algorithm and its variants | Tested only in a small office environment; device diversity was limited (mostly Lenovo tablets, with only five exceptions); only approximately 20 devices were present |
| Tan and Chan (2021) | Use Bayes' theorem to establish the associative probability between bursts of probe requests with selected features; ESPRESSO algorithm | Evaluate performance using discrimination accuracy and V-measure (i.e., determine whether two probe requests originate from the same device but do not count the number of devices) rather than actual occupancy; tested only on probe requests labeled with physical MAC addresses |
| Brik et al. (2008) | Perform a passive analysis of radio frequencies by analyzing the physical layer | Only applicable in controlled lab settings; not suitable for real-world scenarios due to radio interference and the complexity of data collection setup |
| Matte et al. (2016) | Fingerprint probe requests sent by a single device using time-based signature | Not reliable in real-world environments due to scattering problems and multi-path phenomena of timing information |

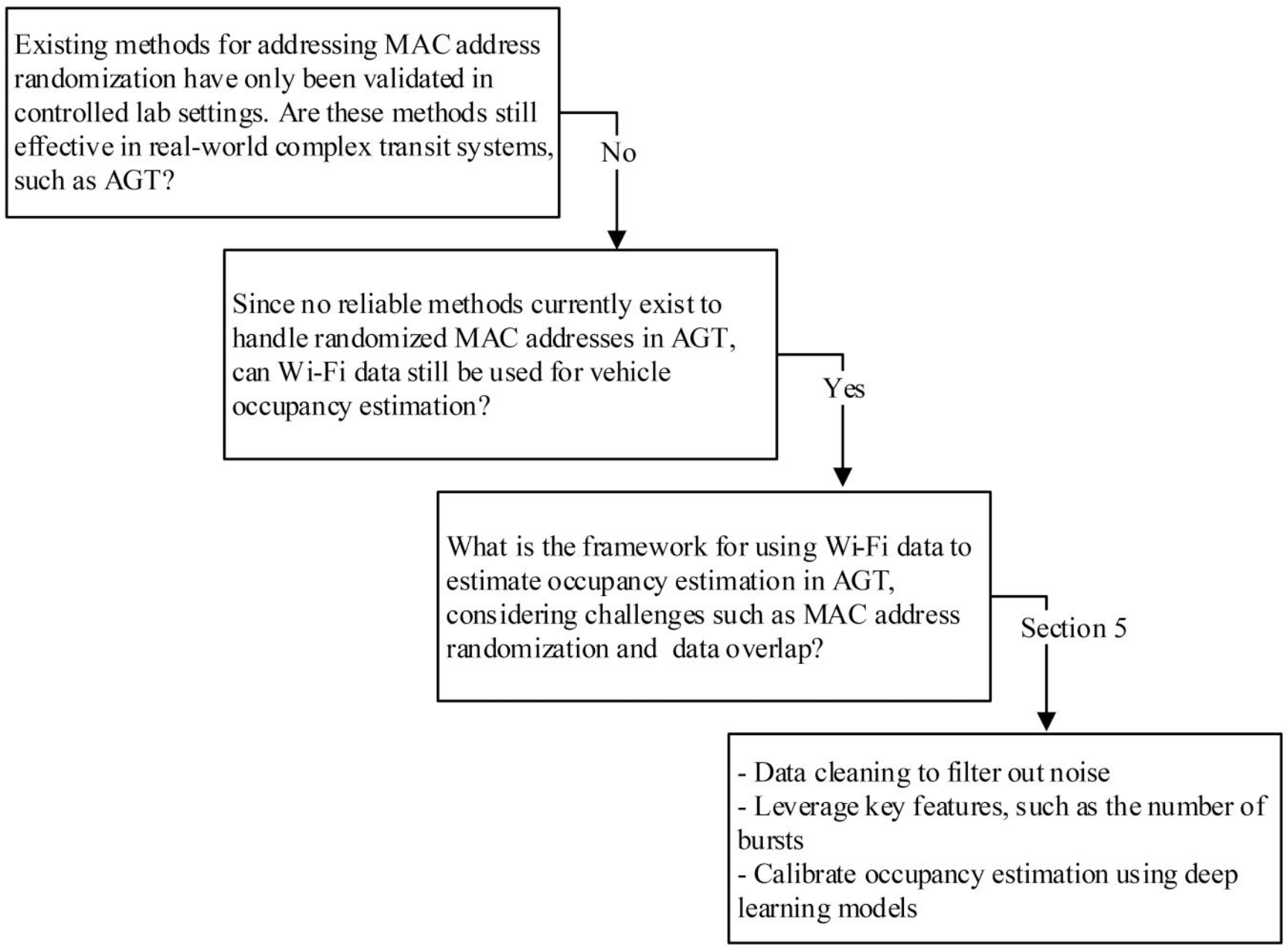

**Figure 3.** Investigating methods for addressing MAC address randomization in AGT.

Furthermore, distinguishing between noise and actual passenger signals poses a significant challenge. While Pu et al. (2021) proposed a fuzzy c-means (FCM) clustering method to address this issue, their approach did not account for MAC address randomization. Consequently, there is a need for further investigation to evaluate the efficacy of fuzzy clustering methods under MAC address randomization.

Moreover, exploring the occupancy estimation models is essential. While traditional machine learning algorithms, such as random forests (RFs), are commonly employed, there is a growing interest in utilizing deep learning methodologies. Pronello et al. (2025) applied Convolutional and Recurrent Neural Networks (CRNN) to address people counting at bus stops, yielding satisfactory results. Chen et al. (2020) utilized Convolutional Deep Bi-directional Long Short-Term Memory (CDBLSTM) for building occupancy estimation models. Additionally, Zhang et al. (2021) investigated various deep learning models for short-term passenger flow forecasting, with their proposed ResLSTM model demonstrating superior performance. These studies highlight the potential of deep learning approaches in improving the accuracy and efficiency of occupancy estimation models.

### 2.4. Summary

Figure 4 summarizes the key challenges in vehicle occupancy estimation using Wi-Fi probe requests. Notably, existing studies that consider MAC address randomization have been conducted primarily in controlled experimental settings, raising concerns about their applicability in complex transit systems like AGT. This study is motivated by the need to validate whether the methods proposed in the literature still perform well in real-world environments and to identify the most effective procedure for such challenging settings.

Based on the literature review, we identify the following limitations:

- Earlier studies on Wi-Fi sniffing system applications in public transit were developed before MAC address randomization was widely adopted, so their methods no longer work today (Section 2.1).
- Recent studies addressing MAC address randomization have been tested only in controlled lab settings with low occupancy and uncrowded environments. Their effectiveness in real-world, high-density transit systems like AGT remains uncertain (Section 2.2).
- Passenger and non-passenger data overlap remains a key challenge. Pu et al. (2021) proposed a fuzzy clustering method to address this issue, but its effectiveness in highly crowded environments and under MAC address randomization remains unknown (Section 2.3).
- Machine learning models are commonly used to calibrate estimation errors in Wi-Fi-based occupancy estimation. While deep learning models hold promise, further investigation is needed to fully understand how they can be effectively applied to this domain (Section 2.3).

Therefore, most studies either lack integration of recent advancements or are limited in scope. A comprehensive and up-to-date methodology is needed to leverage Wi-Fi technology for occupancy estimation in complex real-world transit systems.

## 3. Framework design

The proposed framework is shown in Figure 5. This section describes how Wi-Fi probe requests are collected and pre-processed.

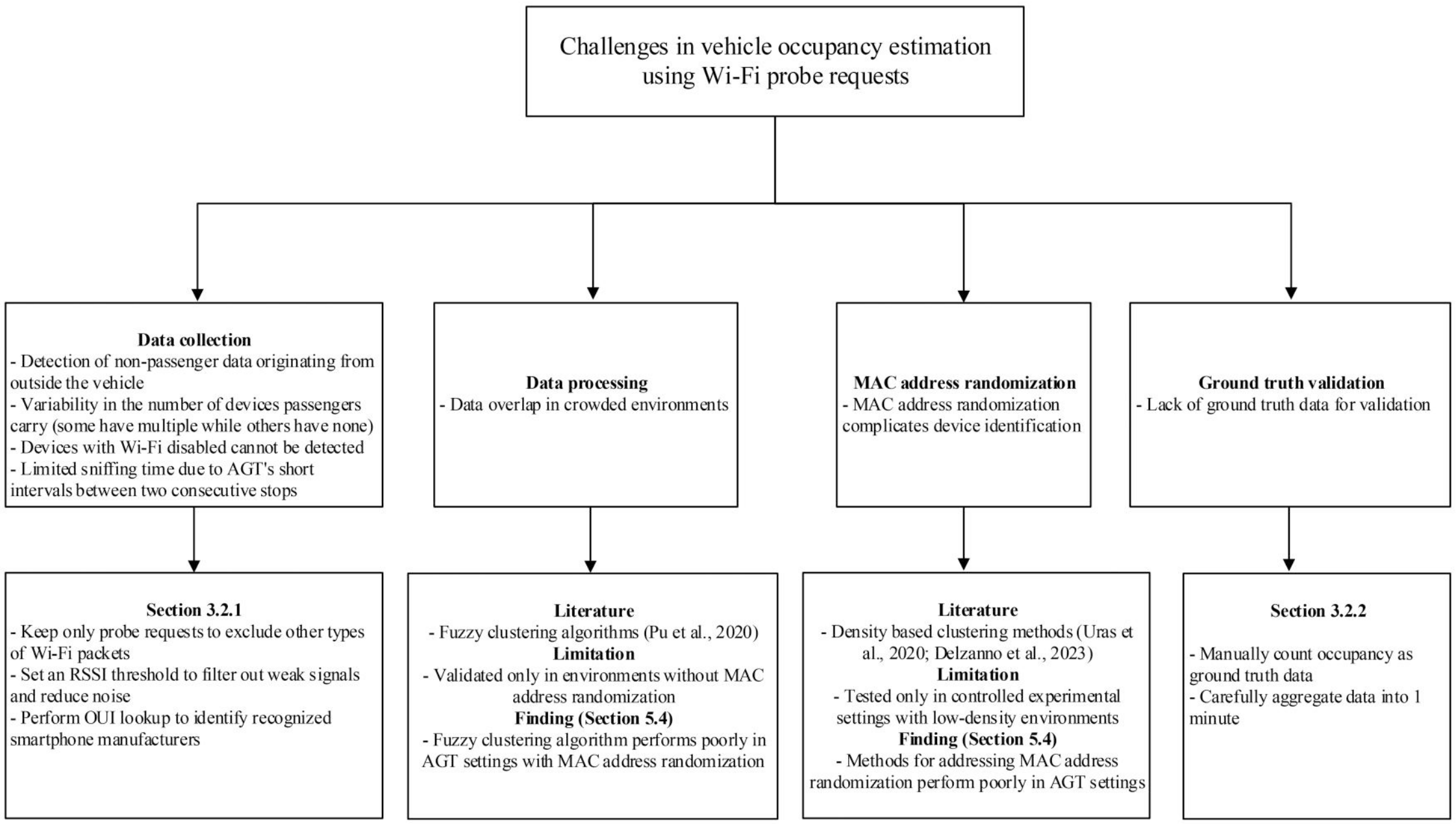

Figure 4. Challenges in vehicle occupancy estimation using Wi-Fi probe requests.

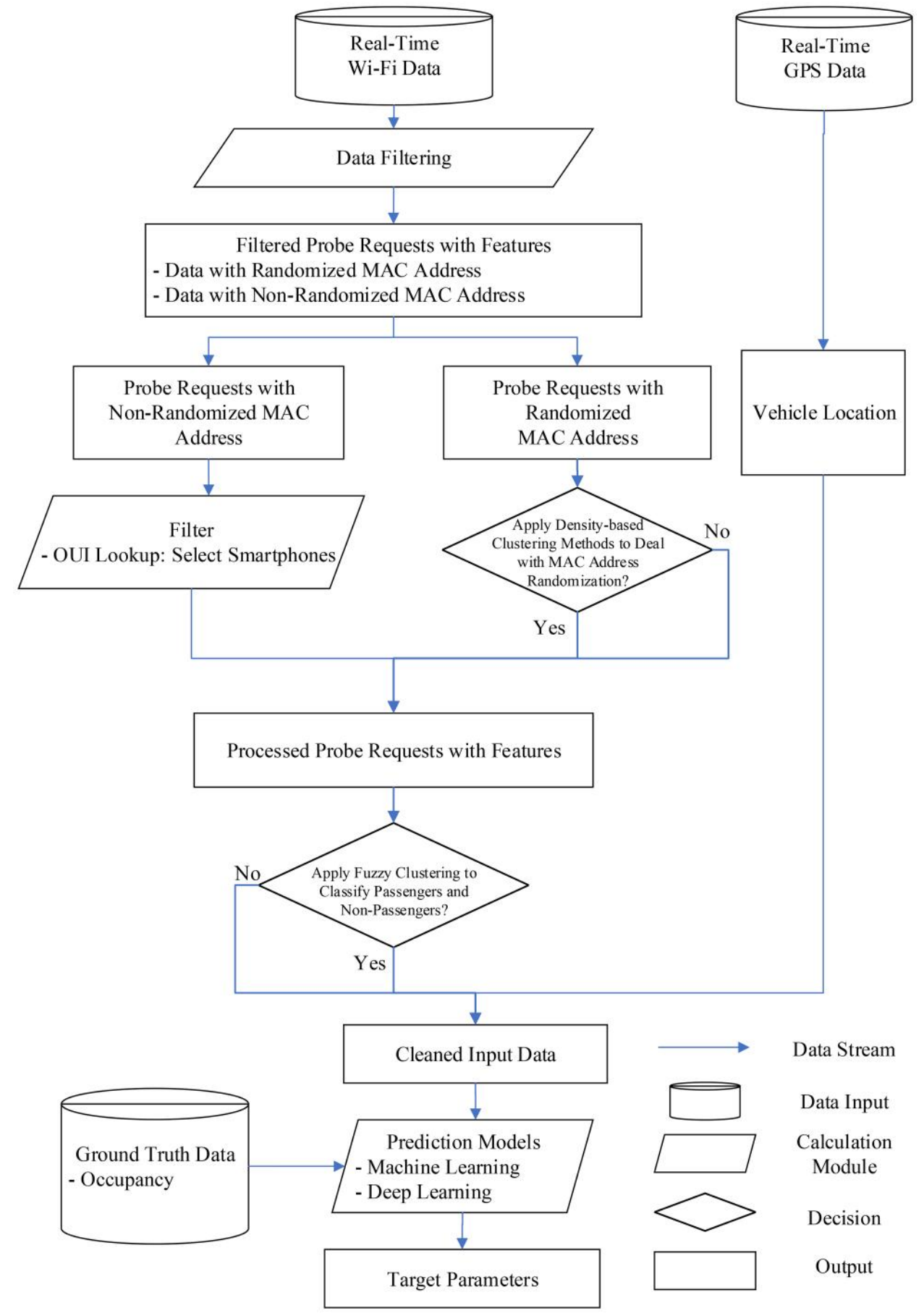

**Figure 5.** Proposed analysis framework.

### 3.1. Hardware setup and data collection

Wi-Fi probe requests are captured by a Wi-Fi sniffer in monitor mode. We employ a USB Wi-Fi adapter TP-Link TL-WN722N that operates within the 2.4–2.4835 GHz frequency range. The adapter is plugged into a single-board computer Le Potato (AML-S905X-CC), which runs Wireshark to save captured Wi-Fi frames. The computer is equipped with a portable monitor and a compact keyboard with a touchpad for monitoring during the data collection, as illustrated in Figure 6. The above hardware setup has flexibility for customization, i.e., it can further incorporate a Bluetooth sensor, a GPS tracker, and a real-time clock depending on specific research needs.

The Wi-Fi sniffer captures all Wi-Fi packets in its range, including those from Wi-Fi devices that are not connected to any Wi-Fi networks. When the sniffer is placed inside a vehicle, it may capture Wi-Fi frames originating from outside the vehicle, as illustrated in Figure 7 (adapted from Surangsrirout and Guo (2024)). Capturing Wi-Fi frames from non-passenger devices, such as those from people in nearby buildings, introduces the problem of potential overestimation (Dunlap et al., 2016; Mehmood et al., 2019). Additionally, other sources of error include not all passengers carrying Wi-Fi-enabled devices, while some carrying multiple devices (Longo et al., 2019). Although these errors cannot be fully eliminated, their impact can be reduced through rigorous data preprocessing and calibration of occupancy estimation using machine learning or deep learning models. These models improve estimation accuracy by learning from historical data, assigning appropriate weights to features, and tuning model parameters to capture the relationship between features and actual occupancy, which help correct systematic biases caused by certain over- or under-estimation patterns.

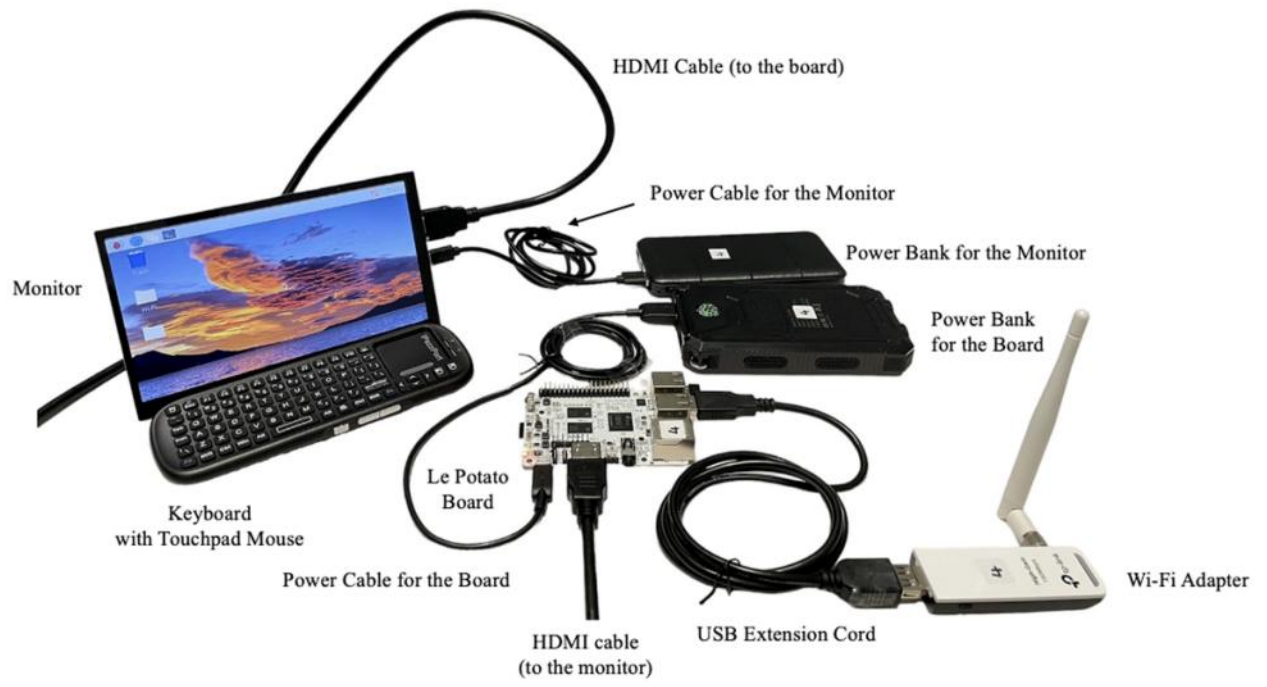

**Figure 6.** Hardware setup.

### 3.2. Data pre-processing and aggregation

#### 3.2.1. Wi-Fi data collection and cleaning

After collecting all Wi-Fi frames, we use the Python package *Pyshark* to extract data from pcapng files.

Before using the data for analysis, potential issues such as missing data and outliers must be addressed, as they can impact accuracy and reliability (Barabino et al., 2017; McLeod, 2007). Our preliminary study indicates that the Wi-Fi sniffer's detection capability is strong enough to capture Wi-Fi data, making it highly unlikely that passenger data are missing due to detection limitations. Instead, the primary challenge lies in filtering out non-passenger data, such as those originating from nearby buildings. To address this, we implement the following data filtering procedure:

- Step 1: Select probe requests only by specifying Wi-Fi frame type as 0 (management frame) and subtype as 4 (probe request).

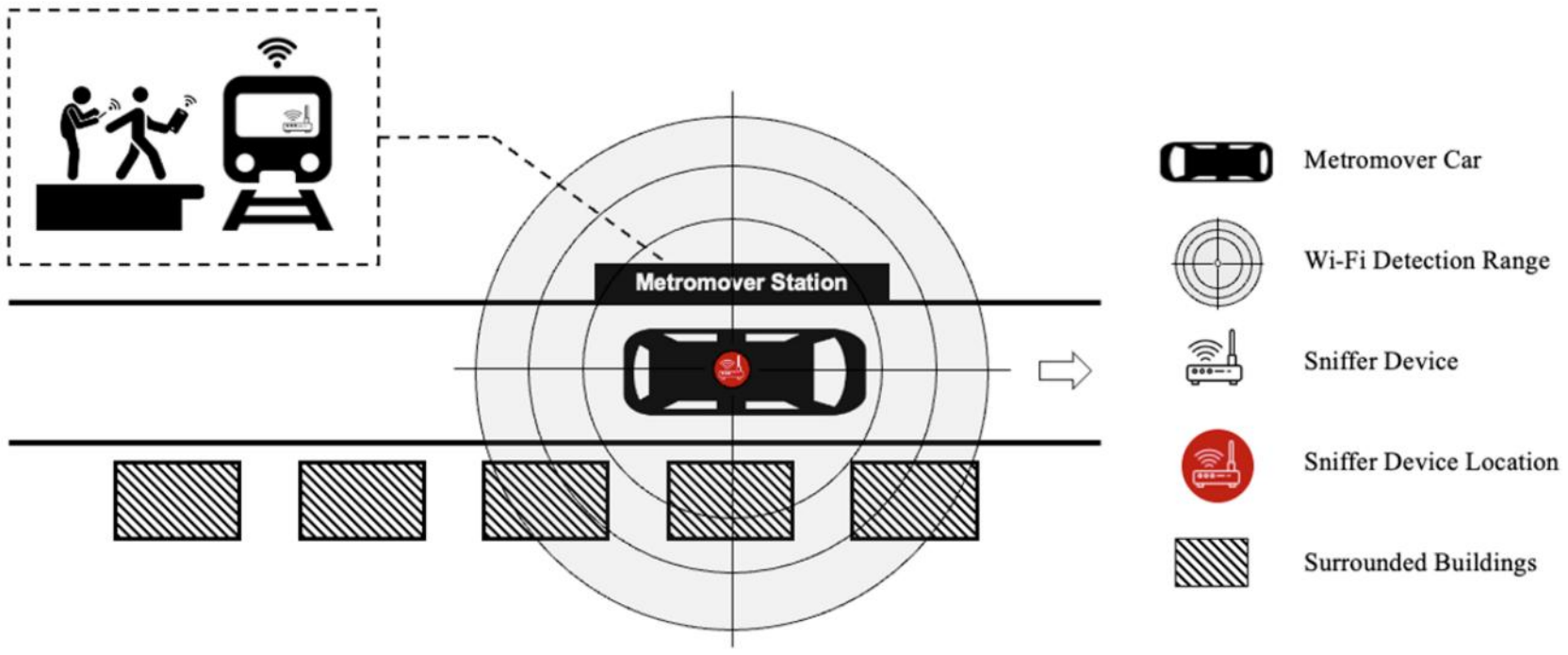

**Figure 7.** Range of Wi-Fi sniffer deployed inside the vehicle.

- Step 2: Choose only those broadcasting probe requests, i.e., with destination address `ff:ff:ff:ff:ff:ff`.
- Step 3: Select unicast probe requests ($b_0 = 0$).
- Step 4: Remove data with RSSI values below −60 dBm.
- Step 5: For non-randomized MAC addresses ($b_1 = 0$), perform an OUI lookup to identify recognized smartphones using the Python package *Ouilookup*.

In Step 2, we filter out probe requests that are not directed to the broadcast address `ff:ff:ff:ff:ff:ff`. According to Bravenec et al. (2023), when the broadcast address is used, the packet is generally directed to all devices on the network, or in the case of wireless communications to all devices in the proximity. In some cases where the packet request is directed to other addresses, it is typically for active scanning and reconnection purposes (IEEE Computer Society LAN/MAN Standards Committee, 2007). For occupancy estimation objectives, we exclude these cases as it is likely that such requests do not originate from passenger devices. Since the Metromover does not provide Wi-Fi access, no passenger device would reconnect.

In Step 4, we set an RSSI threshold due to the high volume of noise in AGT. Based on our preliminary study detailed in Section 5.2, as well as prior research (Delzanno et al., 2023; Oransirikul et al., 2019), probe requests with RSSI below −60 dBm are very likely to originate from non-passenger devices and are thus treated as noise. Unlike Pu et al. (2021), we do not apply fuzzy classification directly because fuzzy algorithms are known for noise sensitivity (Dave, 1993). Instead, fuzzy classification is applied in later stages after most noise has been excluded, as indicated in Figure 5.

### 3.2.2. Passenger data collection and cleaning

Since the Metromover operates as a fare-free system, there is no available AFC data for passenger counts. Instead, vehicle occupancy data were manually recorded. A sample data collection form is depicted in Figure 8.

Several notable scenarios are described as follows. First, the vehicle could become densely crowded, posing challenges for accurate manual passenger counting due to obstructed visibility. Second, the short intervals between stops meant that vehicles could pass through two stations within a minute, complicating data aggregation, as averaging Wi-Fi and passenger data over one minute introduced potential errors. In real-world cases, aggregating data over short intervals has received limited attention. However, addressing this issue is crucial for accurate occupancy estimation. To tackle this, we apply the following data processing techniques:

- If the vehicle was in motion during a specific minute, such as at 2:20 PM, no averaging was necessary; the onboard count of 30 sufficed.
- When the vehicle might be moving or stopped, as observed at 2:21 PM, where it might have not yet reached or might have been waiting at stop 2, the average of the load numbers (30 and 28) yielded an occupancy of 29.
- The more complicated cases arise when the same time appears in two or more rows of the form. For instance, at 2:27 PM, the vehicle might stop at station 5, travel between stations 5 and 6, or remain at station 6. In such cases, we calculate the average of all relevant load numbers within that minute: 38 (upon arrival at station 5), 37 (upon departure from station 5), 37 (upon arrival at station 6), and 61 (upon departure from station 6), resulting the average occupancy 43.25. This approach ensures that changes in occupancy are appropriately captured.

**ON-BOARDPASSENGER CHECK**

**Checker:** Tom **Page:** 4
**Route:** OMNI LOOP **Direction:** Downtown to Omni **Start Station:** Government Center
**Month:** January **Day:** 1 **Year:** 2023 **Start:** 2:19 AM / PM
**Seats:** 4 **Capacity:** 90 **On Board:** 30

| STOP | | | TIME | | PASSENGERS | | | NOTE |
|---|---|---|---|---|---|---|---|---|
| STOP | STATION | TRANSFER | ARR | DEP | ON | OFF | LOAD | |
| 1 | GOVERNMENT CENTER | METRORAIL | 2:19 PM | 2:19 PM | | | 30 | |
| 2 | THIRD STREET | BRICKELL LOOP | 2:21 PM | 2:22 PM | 8 | 10 | 28 | |
| 3 | KNIGHT CENTER | | 2:23 PM | 2:23 PM | 6 | 4 | 30 | |
| 4 | BAYFRONT PARK | | 2:24 PM | 2:25 PM | 8 | 0 | 38 | |
| 5 | FIRST STREET | | 2:26 PM | 2:27 PM | 3 | 4 | 37 | |
| 6 | COLLEGE BAYSIDE | | 2:27 PM | 2:29 PM | 30 | 6 | 61 | |
| 7 | FREEDOM TOWER | | 2:30 PM | 2:30 PM | 3 | 0 | 64 | |
| 8 | PARK WEST | | 2:30 PM | 2:31 PM | 9 | 18 | 55 | |
| 9 | ELEVENTH STREET | | 2:31 PM | 2:31 PM | 6 | 8 | 53 | |
| 10 | MUSEUM PARK | | 2:32 PM | 2:33 PM | 7 | 9 | 51 | |
| 11 | ADRIENNE ARSHT CENTER | | 2:35 PM | 2:35 PM | 8 | 16 | 43 | |
| 12 | SCHOOL BOARD | | 2:35 PM | 2:35 PM | 12 | 14 | 41 | |
| 13 | ADRIENNE ARSHT CENTER | | 2:37 PM | 2:38 PM | 10 | 2 | 49 | |
| 14 | MUSEUM PARK | | 2:40 PM | 2:41 PM | 6 | 6 | 49 | |
| 15 | ELEVENTH STREET | | 2:41 PM | 2:41 PM | 2 | 7 | 44 | |
| 16 | PARK WEST | | 2:41 PM | 2:42 PM | 3 | 0 | 47 | |
| 17 | FREEDOM TOWER | | 2:44 PM | 2:44 PM | 0 | 5 | 42 | |
| 18 | COLLEGE NORTH | INNER LOOP | 2:45 PM | 2:46 PM | 1 | 18 | 25 | |
| 19 | WILKIE D. FERGUSON, JR | BRIGHTLINE | 2:47 PM | 2:47 PM | 1 | 7 | 19 | |
| 20 | GOVERNMENT CENTER | METRORAIL | 2:48 PM | 2:49 PM | 0 | 6 | 13 | |

**Figure 8.** A sample manual count form as the ground truth.

Similar considerations apply to other cases, such as 2:30 PM and 2:35 PM.

Additionally, Figure 8 explains why time-based fingerprinting methods to deal with randomized MAC addresses are unreliable. Consider a passenger traveling from Stop 14 to Stop 16. The vehicle departed Stop 14 at 2:41:06 PM and arrived at Stop 16 at 2:41:53 PM, while the passenger boarded at 2:41:02 PM and alighted at 2:41:57 PM, with a travel time of less than 1 min. Even if the passenger carried a device, it might not have sent a probe request, or only one during the period, making time-based fingerprinting impossible. Therefore, identifying every passenger's device is not feasible.

## 4. Methodology

This section summarizes approaches for occupancy estimation using Wi-Fi technology, including clustering methods for handling randomized MAC addresses, fuzzy clustering for distinguishing passengers from non-passengers, predictive models, and evaluation metrics. The data flow is shown in Table 3. While these methods have shown effectiveness in less crowded environments, their applicability to AGT settings remains uncertain. We first present these methods and later evaluate their performance in AGT environments through our experiments in Section 5.

### 4.1. Clustering methods to deal with randomized MAC address

Clustering methods have been discussed in Section 2.2. Here, we do not aim to recover the true MAC address or identify the device sending the probe request. Instead, researchers cluster probe requests based on feature similarities (Delzanno et al., 2023; Uras et al., 2020). The rationale is that, despite MAC addresses randomization, a device's probe requests should exhibit similar data rates (e.g., a Samsung Galaxy M12 supports higher data rates than an iPhone 6s (Delzanno et al., 2023)) and other attributes.

In real-world scenarios with a large number of devices and extensive data collection, applying the algorithm alone may not suffice. For instance, if the maximum number of probe requests sent by a device within a minute is 30, but there are more than 30 data points within a cluster, overlap occurs. In such cases, it becomes necessary to divide the average probe requests sent by each device within the cluster (Delzanno et al., 2023).

The pseudo-code for the clustering methods to handle randomized MAC address is provided in Algorithm 1. After processing the randomized MAC addresses, each probe request is assigned a representative MAC address, assuming that all data with the same representative MAC address originates from the same device.

**Table 3.** Data flow for different methods in Section 4.

| Section | Method | Objective | Input | Output |
|---|---|---|---|---|
| Section 4.1 | Density-based clustering | Deal with MAC address randomization: map randomized MAC addresses of the same (or similar) device to a single representative address | Probe requests with randomized MAC address | Assigned representative address for randomized MAC address |
| Section 4.2 | Fuzzy clustering | Separate passenger data from non-passenger data | All probe requests | Probe requests from passenger devices, number of passenger devices |
| Section 4.3 | Machine learning/ deep learning | Get occupancy estimation based on number of passengers devices and other features | Number of passenger devices, other features (see Table 6) | Occupancy estimation |
| Section 4.4 | Evaluation metrics | Measure model accuracy | Occupancy estimation and ground truth data | Accuracy |

**Algorithm 1** Clustering methods to handle randomized MAC address

```
Input: Probe requests with randomized MAC address D, clustering
  method M (DBSCAN/DBSCAN_L/OPTICS), MULT/SNGL (MULT: each cluster
  shares multiple devices, SNGL: each cluster is unique to one device),
  parameter Avg (average probe requests sent by each device per
  minute)
Output: Probe requests with assigned representative address for
  randomized MAC address D'
Clustering result C ← Apply clustering method M to data D
for each cluster c ∈ C do
  if SNGL then
    Generate a unique representative MAC address m
    Assign m to all probe requests in cluster c
  else
    num ← number of probe requests in cluster c
    if num ≥ Avg then
      device_count ← round(num ÷ Avg)
    else
      device_count ← 1
    end if
    for i in range(device_count) do
      Generate a unique representative MAC address m
      if i ≠ device_count − 1 then
        Select Avg unassigned probe requests and assign m as their
          identifier
      else
        Assign m to all other unassigned probe requests
      end if
    end for
  end if
end for
```

## 4.2. Passenger and non-passenger classification using fuzzy clustering

Separating probe requests from passenger devices and non-passenger devices can be challenging, so fuzzy logic may be helpful. Pu et al. (2021) utilized the FCM clustering algorithm for this purpose. FCM assigns a degree of membership to each data point, allowing it to belong to multiple clusters. Moreover, Wu et al. (2003) introduced kernel FCM, which uses kernel functions to map data into a higher-dimensional space, enabling the clustering of more complex datasets.

Unlike DBSCAN and OPTICS, which do not require specifying the number of clusters, fuzzy clustering is more adaptable to scenarios where the number of clusters is known. In the context of passenger and non-passenger classification, where the natural clustering involves two groups (passengers and non-passengers) as a priori, fuzzy clustering fits well. However, for clustering methods to deal with randomized MAC addresses, where the number of clusters is unknown, fuzzy clustering is not suitable.

## 4.3. Prediction models

Due to passenger and non-passenger data overlap, the direct output from density-based clustering is not sufficiently accurate. To calibrate occupancy estimation, we apply prediction models, such as machine learning or deep learning, to learn the relationship between the clustering output and actual occupancy.

It is well known that deep learning performs well on large datasets, while machine learning is generally used for smaller datasets. To determine which prediction model is more suitable for the task of vehicle occupancy estimation using Wi-Fi probe requests, both methods were evaluated. Specifically, the following prediction models were tested:

- Machine Learning: To capture the intricate relationships between features, we use ensemble learning methods. Two representative models, RF, and XGBoost Regression (XGB), are selected.
- Deep Learning: Since occupancy levels form a time series, we incorporate RNN/LSTM layers into deep learning models. We select two classic architectures, CRNN and CDBLSTM, which integrate CNN and RNN/LSTM layers. Their architectures are detailed in Pronello et al. (2025); Chen et al. (2020).

We acknowledge that each transit mode has unique features, and no single prediction model consistently outperforms others. This article does not try to propose a new prediction model for AGT but instead establishes foundational insights, showing that deep learning methods consistently outperform traditional machine learning. This highlights the importance of prioritizing deep learning in future prediction model development.

### 4.4. Evaluation metrics

We consider mean absolute error (MAE), root mean square error (RMSE), R-squared score ($R^2$), and mean absolute percentage error (MAPE) as evaluation metrics:

$$MAE = \frac{\sum_{i=1}^{N} \left| \hat{Y}_i - Y_i \right|}{N}, \tag{1}$$

$$RMSE = \sqrt{\frac{\sum_{i=1}^{N} \left( \hat{Y}_i - Y_i \right)^2}{N}}, \tag{2}$$

$$R^2 = 1 - \frac{\sum_{i=1}^{N} \left( Y_i - \hat{Y}_i \right)^2}{\left( Y_i - \overline{Y}_i \right)^2}, \tag{3}$$

$$MAPE = \frac{1}{N} \sum_{i=1}^{N} \frac{\left| \hat{Y}_i - Y_i \right|}{Y_i} \times 100\%, \tag{4}$$

where $\hat{Y}_i$ is the estimated occupancy, $Y_i$ is the ground truth, and $\overline{Y}_i$ is the average of $Y_i$.

Since our data have a higher maximum occupancy, we normalize these metrics to enable comparison with other studies. The normalized metrics are defined as follows:

$$NMAE = \frac{\text{MAE}}{\max_i Y_i} \times 100\%, \tag{5}$$

$$NRMSE = \frac{\text{RMSE}}{\max_i Y_i} \times 100\%. \tag{6}$$

Due to the opaque process of MAC address randomization and the presence of noise, the true mechanism remains unknown and is learned through prediction models. The reliability of the proposed procedure can only be tested using evaluation metrics.

## 5. Experiments and numerical results

### 5.1. An overview of Miami-Dade Metromover

Our pilot study was conducted on the Miami-Dade Metromover's OMNI and BRICKELL loops. The Metromover, an AGT system in downtown Miami, uses electrically powered vehicles on fixed tracks for passenger transport. Key features include its free-ride operation without an AFC system for passenger counting, high-frequency service with cars arriving every 90 s in peak hours, and short travel times between adjacent stops, sometimes passing two stops in a minute. A preliminary study was conducted before the pilot study, with details provided in the following subsection.

### 5.2. Preliminary study

The preliminary study aimed to understand Wi-Fi probing mechanism and MAC address randomization. It also explored the correlation between RSSI and device distance from sniffers, as well as probe request frequency patterns. It included two parts: one in an open space, and the other on a shuttle bus of Florida International University (FIU).

#### 5.2.1. Phase 1: Open space

Phase 1 experiments took place on an empty football field at FIU. We used a Wi-Fi sniffer to detect devices at varying distances. Tests included scenarios with MAC address randomization enabled or disabled and with devices either in use or not.

Initially, we assessed background noise levels, collecting 634 Wi-Fi frames. They predominately had weak signal strength, with 97% having an RSSI below −80 dBm. The collected RSSI values fall into two groups, with the boundary at around −60 dBm.

Subsequently, test devices were positioned at varying distances from the Wi-Fi sniffer, revealing several key findings: (1) Devices within 40 ft (vehicle length) of the sniffer had RSSI stronger than −60 dBm, despite noise frames were detected. Thus, −60 dBm was set as the threshold to filter out noise. (2) Probe requests were often sent in bursts with short intervals between them. (3) Despite MAC address randomization, the MAC address remained consistent within each burst. (4) When in use, devices sent probe requests about every 3 s, while patterns were less clear when not in use.

#### 5.2.2. Phase 2: FIU shuttle experiment

In the second phase, the Wi-Fi sniffer was positioned in the middle of an FIU shuttle, to collect probe data. Data from two complete trips on the same bus route were collected on February 17, 2023, with manual passenger counts serving as ground truth. Each shuttle trip lasted 45 min, with an average of 11,416 Wi-Fi frames collected.

**Table 4.** Data size after each filtration step.

| Line | | Raw data | After Step 1 | After Step 2 | After Step 3 | After Step 4 | After Step 5 |
|---|---|---|---|---|---|---|---|
| OMNI | Training data | 3,280,221 | 204,452 | 199,141 | 199,091 | 117,126 | 107,405 |
| | Test data | 961,998 | 45,146 | 43,524 | 43,511 | 21,397 | 19,598 |
| BRICKELL | Training data | 4,136,054 | 313,797 | 308,978 | 308,946 | 187,668 | 171,911 |
| | Test data | 1,038,604 | 68,927 | 66,974 | 66,971 | 39,777 | 36,630 |

Analysis revealed that about 2.25% of probe requests had non-randomized MAC addresses and weak RSSIs around −80 dBm. These requests were transient and possibly from roadside devices. Most probe requests of our interest had randomized MAC addresses and strong signal strength with RSSIs above −55 dBm. It was again concluded that setting a threshold of −60 dBm effectively filtered out most noise.

## 5.3. Pilot study on Metromover: data collection and summary statistics

Data collection occurred over a two-week period, from March 27 to April 9, 2023, including both weekdays and weekends, with a total of over 100 trips. Daily data collection took place within a four-hour window.

**Table 5.** Proportion of randomized and non-randomized MAC addresses.

| Line | | Non-random MAC (%) | Random MAC (%) | Total (%) |
|---|---|---|---|---|
| OMNI | Train | 2.5 | 97.5 | 100 |
| | Test | 2.6 | 97.4 | 100 |
| BRICKELL | Train | 2.4 | 97.6 | 100 |
| | Test | 2.0 | 98.0 | 100 |

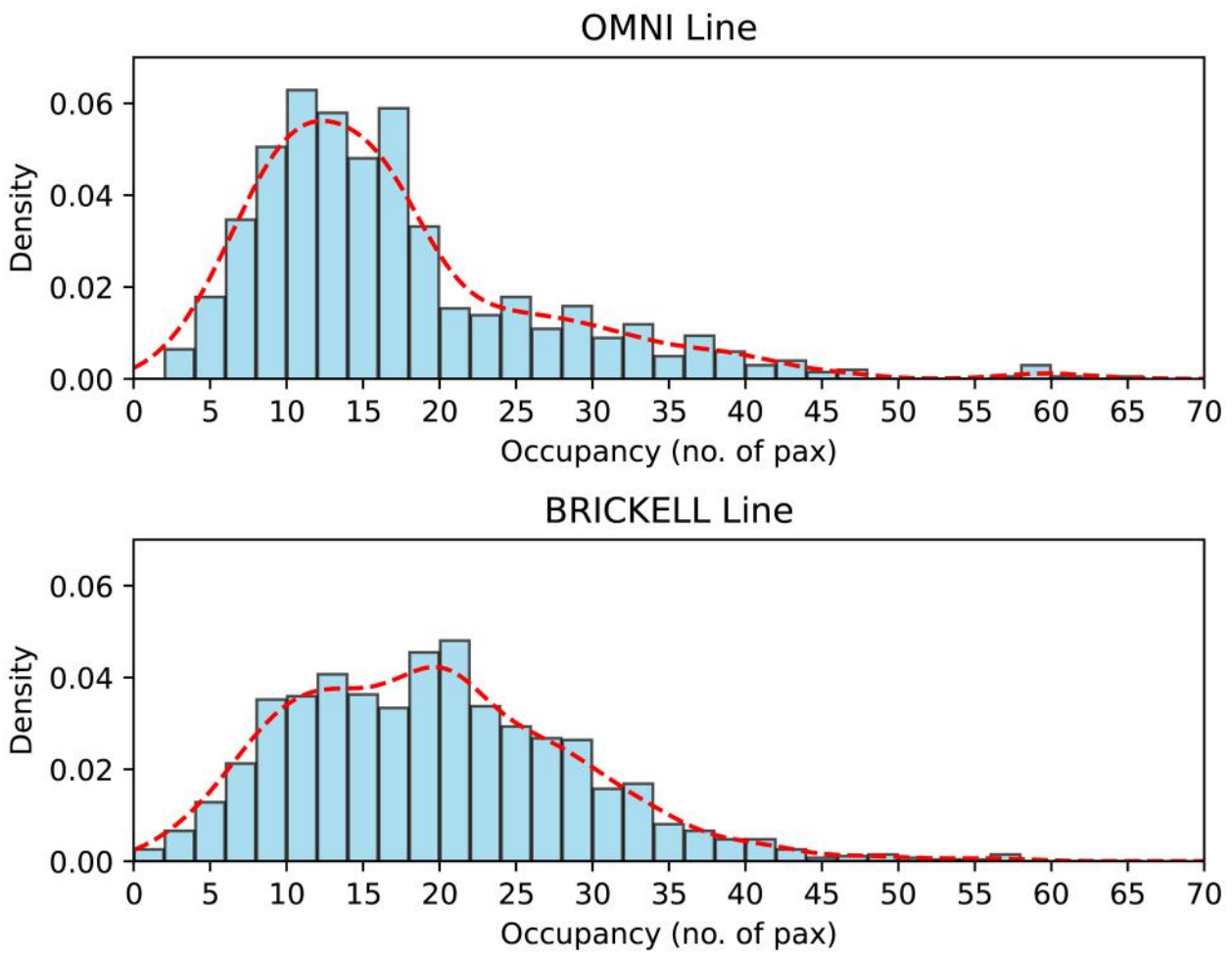

**Figure 9.** Occupancy histogram of training data.

### 5.3.1. Data filtering process

Table 4 displays the data size (the number of Wi-Fi frames) following each stage of the data filtering process in Section 3.2.1. The initial step, which selected probe requests, significantly reduced the data volume. Further filtering in Step 4, which excluded data with RSSI below −60 dBm, further halved the remaining data.

### 5.3.2. Proportion of randomized MAC addresses

For the filtered data, we calculated the proportion of randomized MAC addresses (see Table 5). It is evident that most devices utilized randomized MAC addresses, posing challenges for prediction. The proportion aligns with the FIU shuttle bus preliminary study (2.25% for non-randomized MAC addresses).

### 5.3.3. Distribution of occupancy

Figure 9 shows the occupancy histogram of the training data. Most data points fall within the moderate occupancy range of around 20 passengers. The distribution has a long tail, with some records showing occupancy levels over 50 individuals. Unlike existing research, which typically examines occupancy levels below 35 (Oransirikul & Takada, 2019; Pronello et al., 2025), our study explores higher occupancy levels (up to 64), further validating the proposed methodology in such scenarios.

## 5.4. Occupancy estimation and performance evaluation

### 5.4.1. Features for occupancy estimation

Features we consider are summarized in Table 6, supported by Pu et al. (2021) and Chang et al. (2023). Not all features are incorporated into the predictive model: machine learning models incorporate only those features with the best 5-fold cross-validation performance on the training data. This feature

**Table 6.** Extracted features for occupancy estimation.

| Features | Definition |
|---|---|
| MAC_num | The number of MAC addresses |
| source_random_num | The number of randomized MAC addresses |
| data_rate_mean | The average value of data rate |
| SSID_none_num | The number of probe requests with empty SSID |
| SSID_missing_num | The number of probe requests with missing SSID |
| captured_length_mean | The average value of captured length |
| duration_mean | The average value of duration |
| random_bursts | The number of bursts (with randomized MAC addresses $\geq$ 2 in one minute) |
| RSSI $_{i,j}$ | The number of probe requests with RSSI between $i$ and $j$. E.g., RSSI $_{60,55}$ is the number of probe requests with RSSI between −60 and −55 dBm |
| station_id_$i$ | The indicator of whether the vehicle was at station $i$ during the minute |

**Table 7.** Evaluation of occupancy estimation using machine learning.

| | | | | OMNI | | | | | | BRICKELL | | | | | |
|---|---|---|---|---|---|---|---|---|---|---|---|---|---|---|---|
| | | | | No Fuzzy | | FCM | | Kernel FCM | | No Fuzzy | | FCM | | Kernel FCM | |
| | | | | RF | XGB | RF | XGB | RF | XGB | RF | XGB | RF | XGB | RF | XGB |
| No-clustering | | Training | MAE | 5.05 | 5.09 | 5.69 | 5.51 | 5.71 | 5.54 | 3.91 | 4.21 | 3.94 | 4.27 | 3.98 | 4.19 |
| | | | RMSE | 6.82 | 7.45 | 7.62 | 7.82 | 7.65 | 7.86 | 5.02 | 5.66 | 5.00 | 5.73 | 5.06 | 5.58 |
| | | | MAPE | 0.38 | 0.37 | 0.44 | 0.40 | 0.44 | 0.41 | [a]∞ | ∞ | ∞ | ∞ | ∞ | ∞ |
| | | | $R^2$ | 0.51 | 0.42 | 0.39 | 0.36 | 0.39 | 0.35 | 0.71 | 0.63 | 0.71 | 0.62 | 0.71 | 0.64 |
| | | Test | MAE | 4.42 | 4.54 | 4.53 | 4.65 | 4.46 | 4.67 | 4.43 | 4.45 | 4.45 | 4.44 | 4.52 | 4.36 |
| | | | RMSE | 5.43 | 5.44 | 5.66 | 5.73 | 5.60 | 5.82 | 5.83 | 5.81 | 5.82 | 5.83 | 5.88 | 5.69 |
| | | | MAPE | 0.74 | 0.76 | 0.79 | 0.78 | 0.78 | 0.79 | 0.29 | 0.29 | 0.30 | 0.29 | 0.30 | 0.29 |
| | | | $R^2$ | 0.19 | 0.18 | 0.12 | 0.09 | 0.13 | 0.06 | 0.56 | 0.56 | 0.56 | 0.56 | 0.55 | 0.58 |
| DBSCAN | [b]MULT | Training | MAE | 5.63 | 5.32 | 5.73 | 5.41 | 5.73 | 5.43 | 4.09 | 4.35 | 4.18 | 4.34 | 4.22 | 4.32 |
| | | | RMSE | 7.59 | 7.63 | 7.72 | 7.75 | 7.72 | 7.74 | 5.31 | 5.90 | 5.40 | 5.88 | 5.49 | 5.85 |
| | | | MAPE | 0.43 | 0.39 | 0.44 | 0.39 | 0.44 | 0.39 | ∞ | ∞ | ∞ | ∞ | ∞ | ∞ |
| | | | $R^2$ | 0.40 | 0.39 | 0.38 | 0.37 | 0.38 | 0.37 | 0.68 | 0.60 | 0.67 | 0.60 | 0.66 | 0.61 |
| | | Test | MAE | 4.78 | 4.59 | 4.87 | 4.74 | 4.85 | 4.65 | 4.92 | 4.80 | 5.02 | 4.77 | 5.04 | 4.79 |
| | | | RMSE | 5.78 | 5.52 | 5.94 | 5.81 | 5.92 | 5.73 | 6.36 | 6.23 | 6.50 | 6.23 | 6.50 | 6.24 |
| | | | MAPE | 0.82 | 0.78 | 0.85 | 0.79 | 0.85 | 0.77 | 0.33 | 0.32 | 0.33 | 0.31 | 0.33 | 0.31 |
| | | | $R^2$ | 0.08 | 0.16 | 0.02 | 0.07 | 0.03 | 0.09 | 0.47 | 0.49 | 0.45 | 0.49 | 0.45 | 0.49 |
| | [b]SNGL | Training | MAE | 4.95 | 5.44 | 5.12 | 5.56 | 5.10 | 5.64 | 4.76 | 4.80 | 4.81 | 4.84 | 4.82 | 4.90 |
| | | | RMSE | 6.76 | 7.76 | 6.90 | 7.87 | 6.93 | 7.94 | 6.18 | 6.42 | 6.24 | 6.46 | 6.23 | 6.53 |
| | | | MAPE | 0.37 | 0.40 | 0.39 | 0.41 | 0.39 | 0.41 | ∞ | ∞ | ∞ | ∞ | ∞ | ∞ |
| | | | $R^2$ | 0.52 | 0.37 | 0.50 | 0.35 | 0.50 | 0.34 | 0.56 | 0.53 | 0.55 | 0.52 | 0.56 | 0.51 |
| | | Test | MAE | 5.40 | 4.94 | 5.67 | 5.11 | 5.60 | 5.03 | 5.32 | 5.19 | 5.34 | 5.21 | 5.31 | 5.24 |
| | | | RMSE | 6.45 | 5.99 | 6.79 | 6.26 | 6.74 | 6.21 | 6.87 | 6.65 | 6.90 | 6.68 | 6.88 | 6.67 |
| | | | MAPE | 0.91 | 0.82 | 0.96 | 0.85 | 0.96 | 0.86 | 0.38 | 0.34 | 0.38 | 0.34 | 0.38 | 0.34 |
| | | | $R^2$ | −0.15 | 0.01 | −0.27 | −0.08 | −0.25 | −0.07 | 0.38 | 0.42 | 0.38 | 0.42 | 0.38 | 0.42 |
| DBSCANL | MULT | Training | MAE | 5.68 | 5.16 | 5.19 | 5.60 | 5.31 | 5.54 | 4.15 | 4.49 | 4.28 | 4.49 | 4.19 | 4.42 |
| | | | RMSE | 7.67 | 7.46 | 6.98 | 7.98 | 7.10 | 7.88 | 5.41 | 6.11 | 5.59 | 6.10 | 5.40 | 5.99 |
| | | | MAPE | 0.44 | 0.38 | 0.40 | 0.41 | 0.40 | 0.41 | ∞ | ∞ | ∞ | ∞ | ∞ | ∞ |
| | | | $R^2$ | 0.38 | 0.42 | 0.49 | 0.33 | 0.47 | 0.35 | 0.66 | 0.57 | 0.64 | 0.57 | 0.67 | 0.59 |
| | | Test | MAE | 4.88 | 4.81 | 4.98 | 5.02 | 4.91 | 4.89 | 4.98 | 4.92 | 5.04 | 5.01 | 4.96 | 4.93 |
| | | | RMSE | 5.89 | 5.80 | 6.04 | 6.13 | 5.95 | 5.99 | 6.40 | 6.28 | 6.45 | 6.33 | 6.38 | 6.26 |
| | | | MAPE | 0.83 | 0.77 | 0.81 | 0.89 | 0.81 | 0.84 | 0.32 | 0.32 | 0.33 | 0.32 | 0.33 | 0.32 |
| | | | $R^2$ | 0.04 | 0.07 | −0.01 | −0.04 | 0.02 | 0.01 | 0.46 | 0.49 | 0.46 | 0.48 | 0.47 | 0.49 |
| | SNGL | Training | MAE | 4.99 | 5.37 | 4.84 | 5.35 | 4.86 | 5.43 | 4.70 | 4.92 | 4.71 | 4.93 | 4.67 | 4.95 |
| | | | RMSE | 6.73 | 7.66 | 6.47 | 7.61 | 6.49 | 7.63 | 6.10 | 6.55 | 6.11 | 6.57 | 6.06 | 6.57 |
| | | | MAPE | 0.37 | 0.39 | 0.37 | 0.39 | 0.37 | 0.40 | ∞ | ∞ | ∞ | ∞ | ∞ | ∞ |
| | | | $R^2$ | 0.52 | 0.38 | 0.56 | 0.39 | 0.56 | 0.39 | 0.57 | 0.51 | 0.57 | 0.51 | 0.58 | 0.51 |
| | | Test | MAE | 5.21 | 5.03 | 5.31 | 5.06 | 5.51 | 5.20 | 5.20 | 5.25 | 5.18 | 5.27 | 5.28 | 5.20 |
| | | | RMSE | 6.33 | 6.14 | 6.42 | 6.15 | 6.59 | 6.36 | 6.78 | 6.73 | 6.69 | 6.73 | 6.82 | 6.62 |
| | | | MAPE | 0.87 | 0.82 | 0.88 | 0.81 | 0.91 | 0.84 | 0.37 | 0.34 | 0.36 | 0.35 | 0.37 | 0.36 |
| | | | $R^2$ | −0.11 | −0.04 | −0.14 | −0.04 | −0.20 | −0.12 | 0.40 | 0.41 | 0.41 | 0.41 | 0.39 | 0.43 |
| OPTICS | MULT | Training | MAE | 5.67 | 5.10 | 5.66 | 5.63 | 5.13 | 5.46 | 4.08 | 4.40 | 4.34 | 4.46 | 4.22 | 4.42 |
| | | | RMSE | 7.69 | 7.47 | 7.63 | 7.96 | 6.89 | 7.79 | 5.25 | 5.93 | 5.58 | 6.00 | 5.40 | 5.97 |
| | | | MAPE | 0.44 | 0.37 | 0.44 | 0.41 | 0.40 | 0.40 | ∞ | ∞ | ∞ | ∞ | ∞ | ∞ |
| | | | $R^2$ | 0.38 | 0.42 | 0.39 | 0.34 | 0.50 | 0.36 | 0.68 | 0.60 | 0.64 | 0.59 | 0.67 | 0.59 |
| | | Test | MAE | 4.73 | 4.67 | 5.01 | 4.79 | 4.84 | 4.77 | 4.61 | 4.68 | 4.77 | 4.54 | 4.59 | 4.58 |
| | | | RMSE | 5.71 | 5.64 | 6.10 | 5.83 | 5.85 | 5.81 | 5.99 | 6.05 | 6.25 | 5.91 | 6.02 | 5.94 |
| | | | MAPE | 0.82 | 0.77 | 0.87 | 0.82 | 0.81 | 0.79 | 0.30 | 0.30 | 0.31 | 0.28 | 0.30 | 0.29 |
| | | | $R^2$ | 0.10 | 0.12 | −0.03 | 0.06 | 0.05 | 0.07 | 0.53 | 0.52 | 0.49 | 0.54 | 0.53 | 0.54 |
| | SNGL | Training | MAE | 4.83 | 5.60 | 4.80 | 5.56 | 5.44 | 5.57 | 4.59 | 4.68 | 4.61 | 4.58 | 4.58 | 4.72 |
| | | | RMSE | 6.52 | 7.90 | 6.45 | 7.84 | 7.33 | 7.85 | 5.98 | 6.27 | 6.01 | 6.14 | 5.93 | 6.33 |
| | | | MAPE | 0.36 | 0.42 | 0.36 | 0.40 | 0.41 | 0.41 | ∞ | ∞ | ∞ | ∞ | ∞ | ∞ |
| | | | $R^2$ | 0.55 | 0.35 | 0.56 | 0.35 | 0.44 | 0.35 | 0.59 | 0.55 | 0.59 | 0.57 | 0.60 | 0.54 |
| | | Test | MAE | 4.86 | 4.97 | 5.01 | 4.79 | 5.01 | 4.81 | 4.99 | 5.00 | 5.01 | 4.94 | 4.98 | 5.07 |
| | | | RMSE | 6.06 | 6.00 | 6.25 | 5.94 | 6.21 | 5.95 | 6.53 | 6.44 | 6.55 | 6.39 | 6.50 | 6.61 |
| | | | MAPE | 0.80 | 0.85 | 0.83 | 0.79 | 0.84 | 0.79 | 0.34 | 0.32 | 0.34 | 0.32 | 0.34 | 0.32 |
| | | | $R^2$ | −0.02 | 0.01 | −0.08 | 0.03 | −0.07 | 0.02 | 0.44 | 0.46 | 0.44 | 0.47 | 0.45 | 0.43 |

[a]For the BRICKELL line, a MAPE value of "∞" indicates that the data contain zero occupancy, as shown by Equation (4), where MAPE tends to infinity when $Y_i = 0$.

[b]"MULT" indicates that each cluster shares multiple devices, with the number equaling the size of the cluster divided by the average number of probe requests sent by a smartphone, whereas "SNGL" signifies that each cluster is unique to one device.

selection process ensures that only the most relevant features are used, reducing the risk of overfitting, and improving model robustness. All features are included in deep learning models.

#### 5.4.2. Performance of machine learning algorithms

We investigated two machine learning algorithms, RF and XGB. We also explored clustering methods to deal with randomized MAC addresses and fuzzy

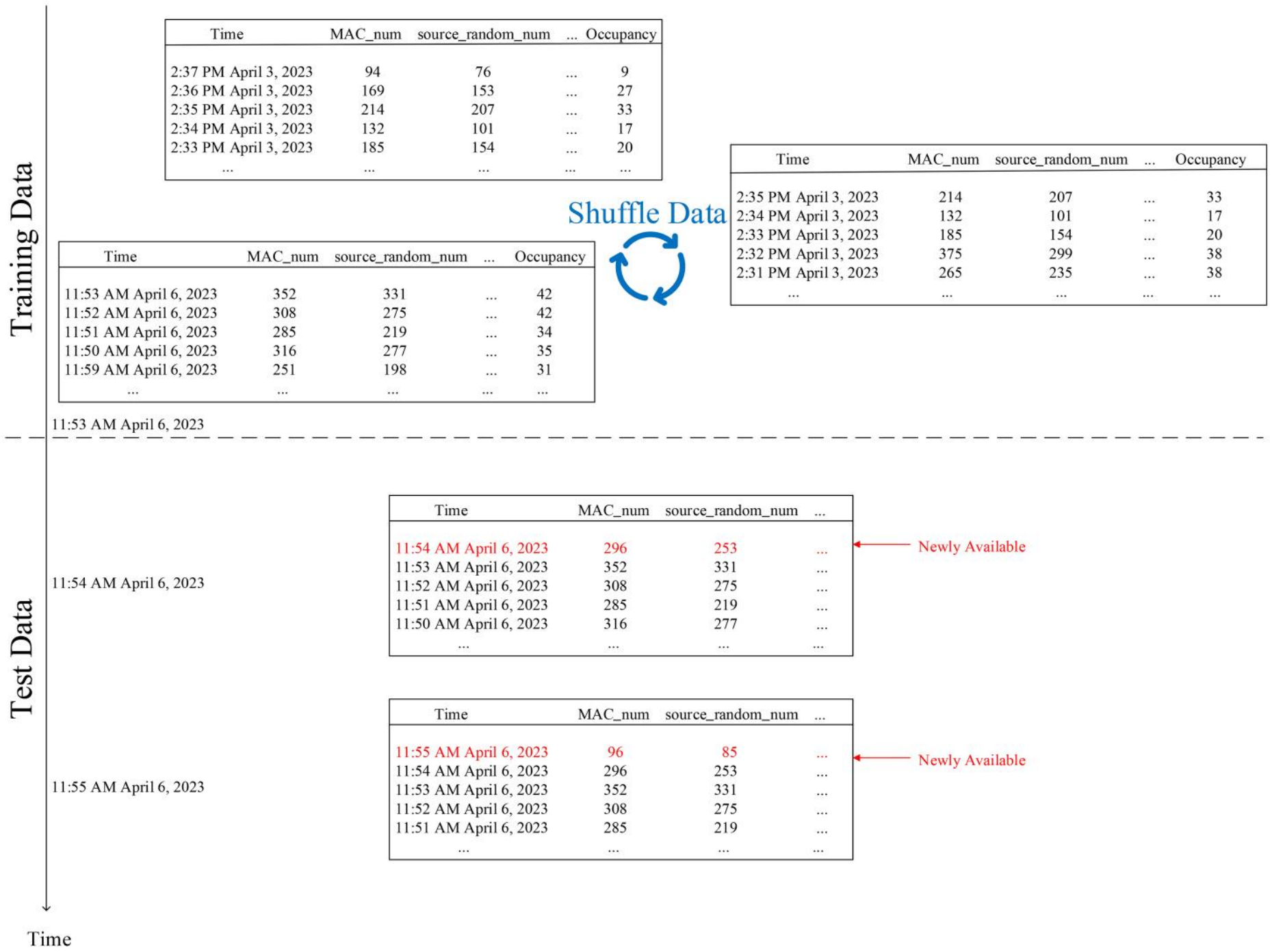

**Figure 10.** Input data of deep learning models.

clustering for passenger and non-passenger separation. The results are presented in Table 7.

- Clustering methods to deal with randomized MAC addresses surprisingly performed worse than no-clustering. Several factors contribute to this outcome. First, the performance of density-based clustering methods heavily depends on the classification capability of features. However, when dealing with large datasets with significant overlaps, clustering MAC addresses solely by these features becomes exceedingly challenging, particularly considering the inherent variability in RSSI values for the same device. Second, the no-clustering approach may benefit from the *burst* concept, where devices send multiple probe requests with consistent MAC address in short intervals, providing valuable information through the feature *random_bursts* (see Kumar & Cunche, 2024). Moreover, when employing any clustering method, two key considerations arise. On the one hand, determining the average number of probe requests sent by smartphones is crucial, particularly for large datasets where clusters may contain probe requests exceeding the maximum possible probe requests sent by a device within one minute. Partitioning each cluster based on the average number of probe requests leads to improved results, as illustrated by the superior performance of "MULT" compared to "SNGL". On the other hand, no single method consistently outperforms the others, underscoring the challenges these clustering methods face when dealing with large datasets.
- Fuzzy clustering for passenger and non-passenger separation unexpectedly showed that the "no fuzzy" option often performs better. Although kernel FCM generally outperforms FCM, both fuzzy methods are worse than "no fuzzy". This is likely due to data overlap. The complex nature of MAC address randomization also compounds the difficulty of passenger and non-passenger separation.
- Machine learning algorithms: In our experiments with no-clustering, RF generally outperforms XGB. However, on the BRICKELL line with kernel FCM, XGB is better. Overall, RF is more effective for its robustness across varying conditions.

In summary, machine learning algorithms demonstrate commendable performance. Despite testing various methods, the "no clustering" and "no fuzzy" scenario consistently outperforms others. Therefore, we focus on this approach in the next section. Since vehicle occupancy data is a time series, incorporating historical data might improve performance. We then test deep learning models using data from the previous $t$ minutes (see Figure 10). For real-time occupancy estimation, probe requests collected over a minute are aggregated into features. These features, along with those from the previous $t$ minutes, are used to estimate occupancy for that minute.

**Table 8.** Evaluation of occupancy estimation using deep learning.

| | | | OMNI | | BRICKELL | |
|---|---|---|---|---|---|---|
| | | | CDBLSTM | CRNN | CDBLSTM | CRNN |
| $t=5$ | Training | MAE | 4.12 | 3.19 | 4.23 | 3.72 |
| | | RMSE | 6.19 | 4.95 | 5.81 | 5.08 |
| | | MAPE | 0.30 | 0.21 | [a]∞ | ∞ |
| | | $R^2$ | 0.60 | 0.74 | 0.61 | 0.71 |
| | Test | MAE | 4.57 | 3.81 | 3.93 | 4.16 |
| | | RMSE | 5.82 | 4.82 | 5.21 | 5.33 |
| | | MAPE | 0.67 | 0.54 | 0.28 | 0.29 |
| | | $R^2$ | 0.07 | 0.36 | 0.65 | 0.63 |
| $t=6$ | Training | MAE | 4.41 | 3.60 | 4.18 | 2.62 |
| | | RMSE | 6.57 | 5.57 | 5.60 | 3.67 |
| | | MAPE | 0.30 | 0.26 | ∞ | ∞ |
| | | $R^2$ | 0.55 | 0.67 | 0.64 | 0.85 |
| | Test | MAE | 3.98 | 3.55 | 3.97 | 4.54 |
| | | RMSE | 5.17 | 4.48 | 5.19 | 5.78 |
| | | MAPE | 0.57 | 0.54 | 0.28 | 0.30 |
| | | $R^2$ | 0.27 | 0.45 | 0.65 | 0.57 |
| $t=7$ | Training | MAE | 2.93 | 2.73 | 4.40 | 3.10 |
| | | RMSE | 3.93 | 4.17 | 5.90 | 4.15 |
| | | MAPE | 0.19 | 0.19 | ∞ | ∞ |
| | | $R^2$ | 0.84 | 0.82 | 0.60 | 0.80 |
| | Test | MAE | 4.44 | 3.86 | 4.00 | 4.62 |
| | | RMSE | 5.50 | 4.96 | 5.15 | 6.12 |
| | | MAPE | 0.61 | 0.52 | 0.30 | 0.32 |
| | | $R^2$ | 0.17 | 0.33 | 0.66 | 0.52 |
| $t=8$ | Training | MAE | 3.94 | 2.93 | 4.54 | 2.76 |
| | | RMSE | 5.87 | 4.59 | 6.00 | 3.77 |
| | | MAPE | 0.31 | 0.22 | ∞ | ∞ |
| | | $R^2$ | 0.64 | 0.78 | 0.59 | 0.84 |
| | Test | MAE | 4.88 | 3.83 | 3.90 | 4.24 |
| | | RMSE | 5.91 | 4.94 | 5.19 | 5.81 |
| | | MAPE | 0.72 | 0.57 | 0.30 | 0.28 |
| | | $R^2$ | 0.05 | 0.33 | 0.65 | 0.56 |
| $t=9$ | Training | MAE | 3.35 | 2.19 | 3.88 | 2.60 |
| | | RMSE | 5.02 | 3.32 | 5.24 | 3.53 |
| | | MAPE | 0.25 | 0.16 | ∞ | ∞ |
| | | $R^2$ | 0.74 | 0.88 | 0.69 | 0.86 |
| | Test | MAE | 4.02 | 3.98 | 3.98 | 4.05 |
| | | RMSE | 4.83 | 5.03 | 5.33 | 5.60 |
| | | MAPE | 0.62 | 0.52 | 0.28 | 0.27 |
| | | $R^2$ | 0.36 | 0.31 | 0.63 | 0.59 |
| $t=10$ | Training | MAE | 3.55 | 2.69 | 4.50 | 2.42 |
| | | RMSE | 5.27 | 4.38 | 5.91 | 3.29 |
| | | MAPE | 0.25 | 0.19 | ∞ | ∞ |
| | | $R^2$ | 0.71 | 0.80 | 0.60 | 0.88 |
| | Test | MAE | 3.77 | 3.93 | 3.94 | 4.24 |
| | | RMSE | 4.58 | 4.90 | 5.12 | 5.70 |
| | | MAPE | 0.56 | 0.56 | 0.30 | 0.30 |
| | | $R^2$ | 0.43 | 0.35 | 0.66 | 0.58 |
| $t=11$ | Training | MAE | 3.84 | 2.67 | 4.25 | 2.00 |
| | | RMSE | 5.74 | 3.95 | 5.86 | 2.70 |
| | | MAPE | 0.26 | 0.18 | ∞ | ∞ |
| | | $R^2$ | 0.66 | 0.84 | 0.61 | 0.92 |
| | Test | MAE | 4.33 | 3.71 | 4.07 | 4.50 |
| | | RMSE | 5.17 | 4.62 | 5.35 | 5.93 |
| | | MAPE | 0.63 | 0.51 | 0.27 | 0.29 |
| | | $R^2$ | 0.28 | 0.42 | 0.63 | 0.55 |

[a]For the BRICKELL line, a MAPE value of "∞" indicates that the data contain zero occupancy, as shown by Equation (4), where MAPE tends to infinity when $Y_i = 0$.

### 5.4.3. Performance of deep learning algorithms

We tested two deep learning frameworks, CRNN and CDBLSTM. The results are presented in Table 8.

In contrast to machine learning methods, deep learning models show significant performance improvements. The CDBLSTM model performs best on the OMNI line with a time lag of $t = 10$, and on the BRICKELL line with $t = 8$. Similarly, the CRNN model performs best with $t = 6$ on the OMNI line and $t = 9$ on the BRICKELL line. While CRNN outperforms CDBLSTM on the OMNI line, the reverse is true for the BRICKELL line. This discrepancy may stem from various factors, including line-specific attributes, and challenges in model training, such as overfitting or underfitting. Overall, deep learning models outperform machine learning approaches, primarily due to their ability to leverage historical data and advanced neural network capabilities.

**Table 9.** Summary of performance on test data using different models.

| | OMNI | | | | BRICKELL | | | |
|---|---|---|---|---|---|---|---|---|
| | RF | XGB | CDBLSTM | CRNN | RF | XGB | CDBLSTM | CRNN |
| MAE | 4.42 | 4.54 | 3.77 | 3.55 | 4.43 | 4.45 | 3.90 | 4.05 |
| RMSE | 5.43 | 5.44 | 4.58 | 4.48 | 5.83 | 5.81 | 5.19 | 5.60 |
| MAPE | 0.74 | 0.76 | 0.56 | 0.54 | 0.29 | 0.29 | 0.30 | 0.27 |
| $R^2$ | 0.19 | 0.18 | 0.43 | 0.45 | 0.56 | 0.56 | 0.65 | 0.59 |

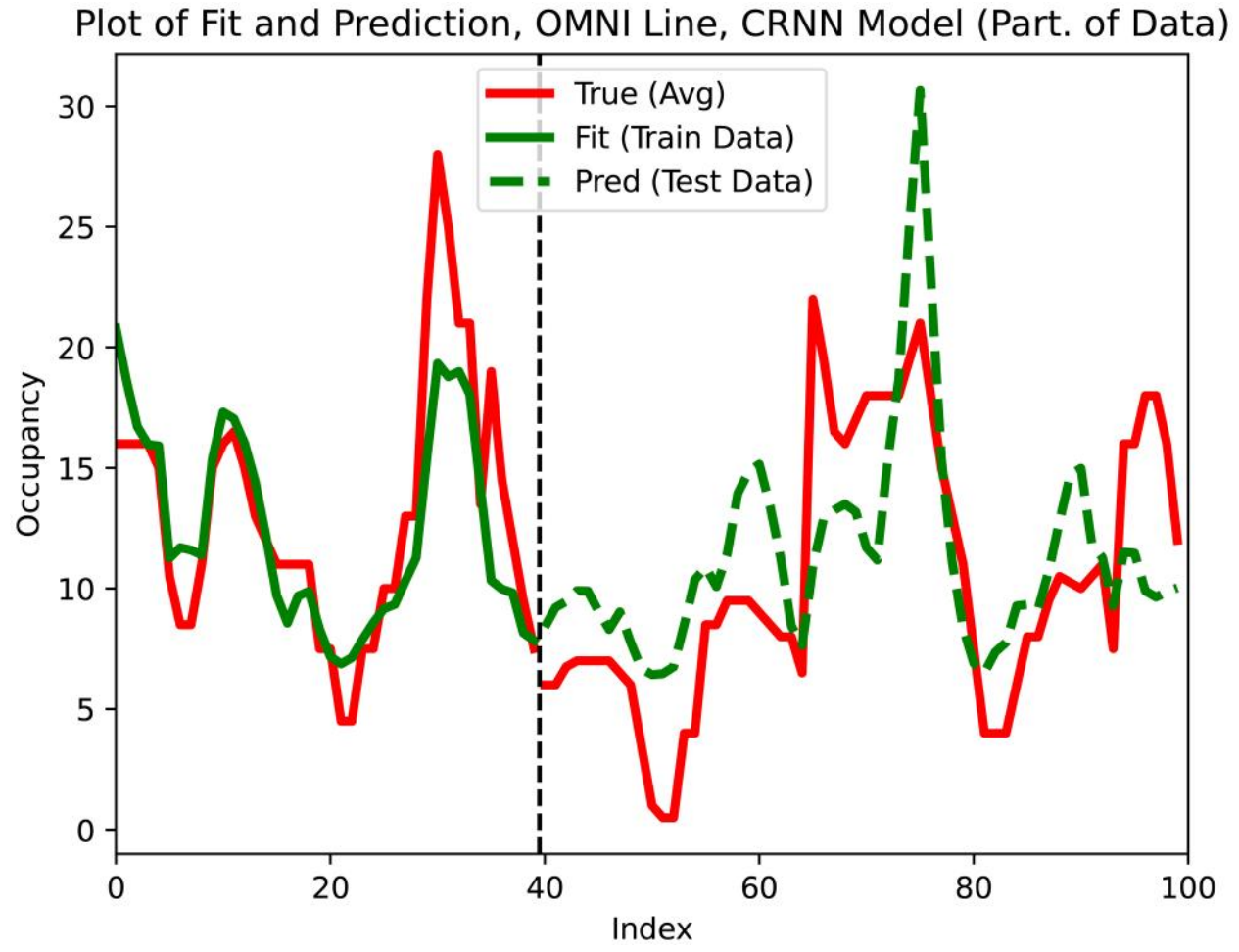

**Figure 11.** Figure depicting the best model performance (CRNN, $t=6$) on a subset of the OMNI line data.

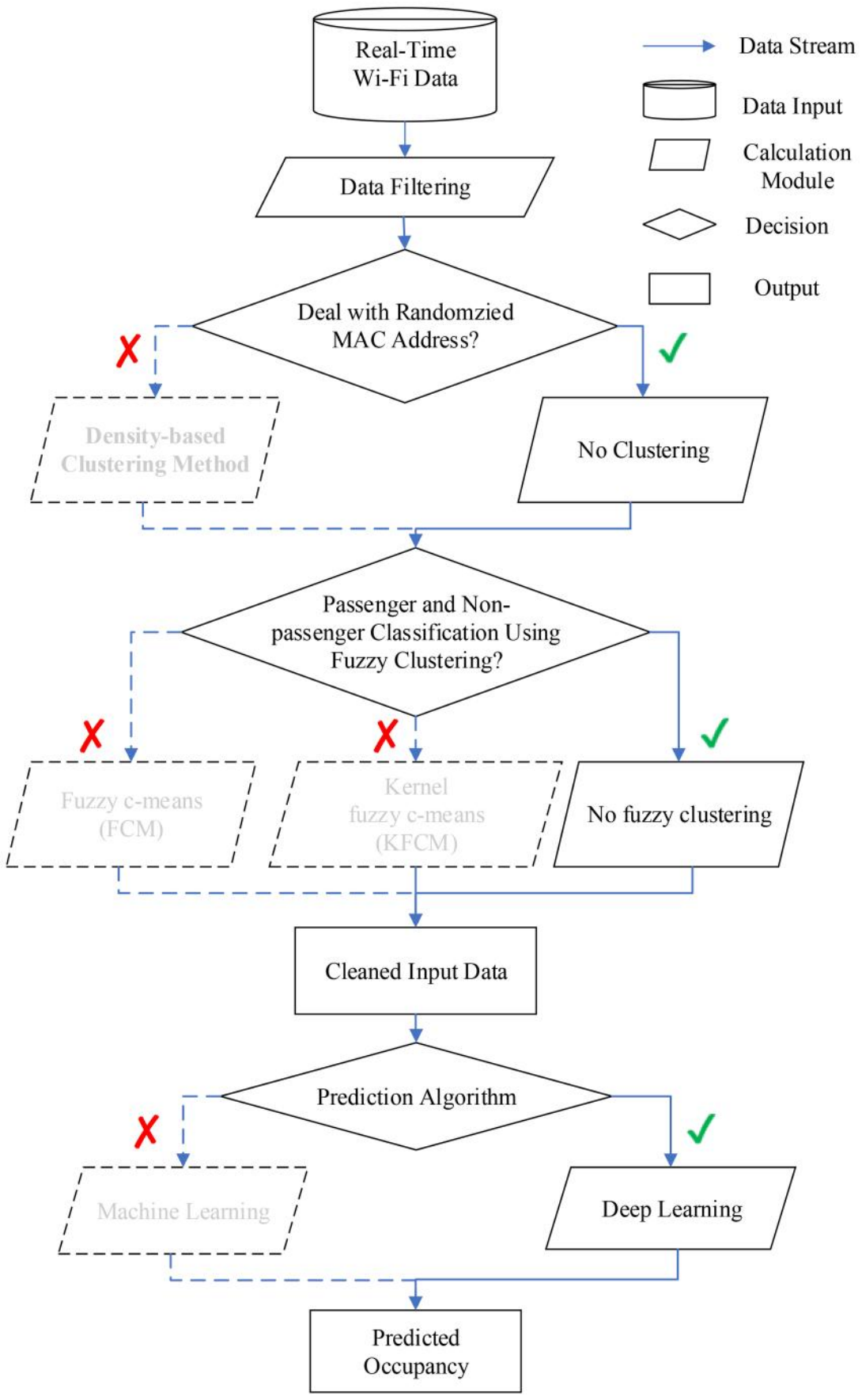

**Figure 12.** Flowchart depicting experimental results.

**Table 10.** Comparison of performance with other studies.

| | | | Reported value | | Normed value | |
|---|---|---|---|---|---|---|
| | Max occupancy | Randomized MAC | MAE | RMSE | NMAE (%) | NRMSE (%) |
| Oransirikul and Takada (2019) | 30 | ✗ | 4.16 | 5.89 | 13.87 | 19.63 |
| Pu et al. (2021) | 40 | ✗ | 2.08 | 3.82 | 5.20 | 9.56 |
| Our results | 64 | ✓ | 3.55 | 4.48 | 5.55 | 7.00 |

### 5.5. Discussion

Table 9 summarizes the performance of different prediction models on the test data, reaffirming that deep learning models consistently outperform machine learning models.

For additional validation of deep learning models, we plot a subset of the estimation performance of the best model, CRNN with $t = 6$ for the OMNI line, as depicted in Figure 11. The figure showcases the model's ability to accurately predict occupancy trends, validating the use of deep learning for occupancy estimation.

Figure 12 outlines the key procedures that lead to superior performance. "No clustering" and "no fuzzy" yield the best results compared to various approaches. Additionally, deep learning outperforms machine learning in the task.

Table 10 shows that our normed metrics match those of other studies. It is also important to note that we are addressing unique challenges and sources of error, including:

- Inaccurate ground truth data: The data collection accuracy becomes unreliable with increasing passenger volumes, and vehicles passing two stations within a minute cause errors and data loss during the data aggregation process.
- MAC address randomization: Currently, there is no reliable, up-to-date method to physically de-randomize MAC addresses. Therefore, we utilize alternative features for occupancy prediction, such as the number of bursts.
- Difficulty in distinguishing passenger data: Lab-controlled experiments may separate non-passenger data, but real-world complex systems like AGT introduce more data and noise, complicating the task. Fuzzy clustering is not effective here.

### 5.6. Application for other regions

The methodology developed in this study was tested on AGT, one of the most complex transit systems. Its unique challenges, short travel times between stops, high occupancy, and a crowded environment, make Wi-Fi-based occupancy estimation particularly difficult due to data overlap and limited time to capture data from passenger devices. Given these challenges, applying Wi-Fi technology to AGT is more challenging than in other transit systems. Since the methodology performs well in this complex setting, we believe it can be effectively applied to other transit systems with fewer data overlap issues and longer data collection windows.

Wi-Fi-based vehicle occupancy estimation can be easily extended to other regions. First, the Wi-Fi sniffer must be installed in the vehicle. Then, the transit agency should collect enough Wi-Fi data and corresponding ground truth occupancy data to train the deep learning model. Once trained, the system can estimate real-time occupancy.

## 6. Conclusion

This study represents an innovative endeavor in leveraging Wi-Fi probe requests to estimate vehicle occupancy in AGT systems. We developed a comprehensive framework, from data collection to prediction models. We tested different approaches in a pilot study on the Miami-Dade Metromover system. Two key findings emerged: first, there are currently no reliable methods to deal with randomized MAC addresses or to separate passenger data from non-passengers in real-world transit systems; second, deep learning models significantly outperform machine learning models. Our results match existing literature in normed metrics like NMAE and NRMSE despite being based on a significantly larger sample dataset. Our results prove the validity of occupancy estimation using Wi-Fi in complex settings even with MAC address randomization. This study offers valuable insights for transit agencies, supporting decision-making and improving service quality in the following ways: (1) Vehicle occupancy estimation using Wi-Fi probe requests is a cost-effective solution applicable to nearly all transit modes and achieves commendable accuracy through deep learning models. (2) By analyzing occupancy patterns, agencies can optimize schedules and vehicle dispatching to reduce overcrowding and improve passenger comfort. (3) Occupancy information can be further integrated into transit apps to enhance the passenger experience by providing real-time crowding data, enabling riders to make informed travel choices.

Despite the promising results, the complexities of real-world scenarios necessitate further research into new feature extraction methods. Additionally, refining and exploring deep learning models could further enhance model accuracy. Incorporating real-time GPS data might further improve estimation performance (Pu et al., 2021). This aspect was not covered in this study but will be considered in future research. Furthermore, each occupancy estimation method has its advantages and limitations. To enhance estimation accuracy and robustness, transit agencies may consider integrating multiple data sources for calibration, as demonstrated by Dib et al. (2023).

## Acknowledgments

Comments from anonymous reviewers and editors have significantly improved this article. The authors are grateful for their assistance. The authors are also grateful to Doug Bermudez from Miami-Dade County DTPW, Dr. Priyanka Alluri, and Dr. Albert Gan from Florida International University for their valuable support in data collection. The authors would like to thank Ms. Nattakarn Surangsrirout of Florida International University for providing multiple figures related to the Wi-Fi sniffer discussed in the article.

## Disclosure statement

No potential conflict of interest was reported by the author(s).

## Funding

Part of this research was funded by the Florida Department of Transportation (FDOT) Public Transit Office.

## Data availability statement

The data that support the findings of this study are available on request from the corresponding author.